\documentclass[12pt]{article}
\usepackage{amsmath,amssymb,epsfig}
\usepackage{graphicx,floatflt,subfigure}
\usepackage{cite}

\makeatletter

\renewcommand\section{\@startsection {section}{1}{\z@}%
                                   {-3.5ex \@plus -1ex \@minus -.2ex}%
                                   {2.3ex \@plus.2ex}%
                                   {\normalfont\large\bfseries}}

\renewcommand\subsection{\@startsection{subsection}{2}{\z@}%
                                     {-3.25ex\@plus -1ex \@minus -.2ex}%
                                     {1.5ex \@plus .2ex}%
                                     {\normalfont\normalsize\bfseries}}

\def\unit{{1\kern-.65ex {\rm l}}}
\def\1{{1\kern-.65ex {\rm l}}}

\newcount\hour \newcount\minute
\hour=\time \divide \hour by 60
\minute=\time
\def\now{%
\ifnum \hour<13
  \ifnum \hour=0 \advance \hour by 12 \number\hour:\else \number\hour:\fi%
     \ifnum \minute<10 0\fi%
     \number\minute%
\ A.M.%
\else \advance \hour by -12 \number\hour:%
  \ifnum \minute<10 0\fi%
  \number\minute%
  \ P.M.%
\fi%
}
\makeatother

\newcommand{\be}{\begin{equation}}
\newcommand{\ee}{\end{equation}}
\newcommand{\ba}{\begin{aligned}}
\newcommand{\ea}{\end{aligned}}

\def\m1{\left(-1\right)^{F_i}}

\makeatletter
\def\sla@#1#2#3#4#5{{%
  \setbox\z@\hbox{$\m@th#4#5$}%
  \setbox\tw@\hbox{$\m@th#4#1$}%
  \dimen4\wd\ifdim\wd\z@<\wd\tw@\tw@\else\z@\fi
  \dimen@\ht\tw@
  \advance\dimen@-\dp\tw@
  \advance\dimen@-\ht\z@
  \advance\dimen@\dp\z@
  \divide\dimen@\tw@
  \advance\dimen@-#3\ht\tw@
  \advance\dimen@-#3\dp\tw@
  \dimen@ii#2\wd\z@  \raise-\dimen@\hbox to\dimen4{%
    \hss\kern\dimen@ii\box\tw@\kern-\dimen@ii\hss}%
  \llap{\hbox to\dimen4{\hss\box\z@\hss}}}}
\def\slashed#1{%
  \expandafter\ifx\csname sla@\string#1\endcsname\relax
    {\mathpalette{\sla@/00}{#1}}%
  \else
    \csname sla@\string#1\endcsname
  \fi}
\makeatother

\begin{document}

\baselineskip=18pt  
\numberwithin{equation}{section}  
\allowdisplaybreaks  



%
%


\thispagestyle{empty}

\vspace*{-2cm}
\begin{flushright}
{\tt arXiv:0808.1571[hep-th]}\\
CALT-68-2691\\
\end{flushright}

\vspace*{1.2cm}
\begin{center}
 {\LARGE Gauge Mediation in F-Theory GUT Models\\}
 \vspace*{1.5cm}
 Joseph Marsano, Natalia Saulina and Sakura Sch\"afer-Nameki\\
 \vspace*{.8cm}
{\it California Institute of Technology}\\
{\it 1200 E California Blvd, Pasadena, CA 91125, USA}\\[1ex]

{\tt marsano, saulina, ss299 @theory.caltech.edu}

 \vspace*{0.8cm}
\end{center}
\vspace*{.5cm}

\noindent
We study a simple framework for gauge mediated supersymmetry-breaking in local GUT models based on $F$-theory 7-branes and demonstrate that a mechanism for solving both the $\mu$ and $\mu/B_{\mu}$ problems emerges in a natural way.  In particular, a straightforward coupling of the messengers to the Higgs sector leads to a geometry which not only provides us with an approximate $U(1)_{PQ}$ symmetry that forbids the generation of $\mu$ at the GUT scale, it also forces the SUSY-breaking spurion field to carry a nontrivial PQ charge.
This connects the breaking of SUSY to the generation of $\mu$ so that the same scale enters both.  Moreover, the messenger sector naturally realizes the D3-instanton triggered SUSY-breaking model of \cite{ourinst} so this scale is exponentially suppressed relative to $M_{GUT}$.  The effective action at low scales is in fact precisely of the form of the "sweet spot supersymmetry" scenario studied by Ibe and Kitano in \cite{Ibe:2007km}.

\newpage
\setcounter{page}{1} 



\tableofcontents


\section{Introduction}\label{sec:intro}

Due to the large separation between the Planck and electroweak scales, producing realistic models of particle physics from string compactifications has proven to be a daunting challenge.
This task can be somewhat simplified, however, by noting that if one introduces gauge degrees of freedom using D-branes, the particles that are observed at accelerators are inextricably bound to the branes and hence, at sufficiently low energies, do not probe the full compactified geometry.  This has led several groups to advocate a bottom-up approach to model building in string theory, where one studies local geometries which capture only the structure relevant for particle physics \cite{Aldazabal:2000sa,Gray:2006jb,Verlinde:2005jr,Donagi:2008ca,Beasley:2008dc,Beasley:2008kw}.

A particularly optimistic scenario for the success of bottom-up model building in string theory is that of gauge mediated supersymmetry breaking because, in this case, both the visible and hidden sectors as well as their mediation can be captured within a single effective field theory decoupled from gravity.  This would allow for the possibility that a single local construction in string theory could describe all of the essential physics of supersymmetry-breaking.  Models of this type are also well-motivated from a phenomenological point of view as the flavor-blindness of gauge mediation alleviates potential conflicts with the current bounds on FCNCs.

In this paper, we seek to study supersymmetry-breaking and its mediation to the visible sector within a framework that holds great promise for realistic model building in string theory, namely that of intersecting 7-branes described by local $F$-theory "compactifications" \cite{Donagi:2008ca,Beasley:2008dc,Beasley:2008kw}.  See also \cite{Hayashi:2008ba,Aparicio:2008wh,Buchbinder:2008at} for related work in this direction.  As described by Beasley, Heckman, and Vafa (BHV) in \cite{Beasley:2008dc,Beasley:2008kw}, one can successfully engineer a wide variety of supersymmetric GUTs using collections of intersecting 7-branes that are described in $F$-theory by a certain class of local Calabi-Yau 4-folds.  In their setup, the visible sector gauge group is housed on a single stack of 7-branes which wraps a compact 4-cycle and matter is introduced either by breaking the gauge group with fluxes or intersecting the stack with additional "matter branes".  As emphasized in \cite{Beasley:2008kw}, worldvolume fluxes also provide a natural way to break the GUT gauge group to that of the MSSM.

A simple way to incorporate gauge mediated supersymmetry breaking into this framework has in fact already been suggested by BHV in \cite{Beasley:2008kw}.  In this approach, one engineers a pair of messenger fields, $f$ and $\bar{f}$, in the $\mathbf{5}$ and $\mathbf{\overline{5}}$ of $SU(5)$ which couple to an additional GUT singlet field $X$.  By assuming that $X$ obtains an $F$-component expectation value from new physics away from the GUT brane, one then has a simple model of gauge mediation.  Quite nicely, the construction by which one obtains these messenger fields is precisely what was used in \cite{ourinst} to build a Polonyi model in which supersymmetry breaking is triggered by a D3-instanton.  Thus, we get SUSY-breaking quite naturally in this framework.  The use of stringy instantons to generate small parameters needed for particle physics, including SUSY-breaking parameters as well as $\mu$-terms and neutrino masses, is of course not new and has been considered before in various contexts by a number of groups \cite{Blumenhagen:2006xt,Ibanez:2006da,Buican:2006sn,Cvetic:2007ku,Argurio:2007qk,Antusch:2007jd,Ibanez:2007tu,Cvetic:2008hi}.  
Previous studies of gauge mediation in string theory include \cite{Diaconescu:2005pc,GarciaEtxebarria:2006rw,Kawano:2007ru, Floratos:2006hs, Cvetic:2007qj, Buican:2008qe, Cvetic:2008mh, Kumar:2008cm}.

Any discussion of gauge mediated models, however, must also come to grips with the $\mu$ and $\mu/B_{\mu}$ problems (see e.g. \cite{Giudice:1998bp} for a review).  A common mechanism for explaining the relatively small size of $\mu$ is to build a model with an approximate $U(1)_{PQ}$ symmetry that forbids it and then add some dynamics into the model which breaks this symmetry at a lower scale.  Alternatively, however, one can try to instead arrange for the SUSY-breaking field $X$ to carry $PQ$ charge.  In that case, the same instanton which breaks supersymmetry also triggers the breaking of $U(1)_{PQ}$ and consequently $\mu$ is naturally generated at a scale comparable to the soft mass parameters.  By contrast, $B_{\mu}$ remains forbidden so it is identically zero until RG running of the MSSM kicks in below the messenger scale and generates it.  This approach has been studied in great detail in the so-called "sweet spot supersymmetry" scenario
of Ibe and Kitano
\cite{Ibe:2007km,Ibe:2007gf,Ibe:2007mr} who demonstrated that models of this type can have very favorable phenomenology when the Higgs and SUSY-breaking sectors are coupled at the GUT scale and the gravitino mass sits at the $1$ GeV "sweet spot".  This idea has also been incorporated into a GUT model \cite{Kitano:2006wm} in which SUSY-breaking is triggered by a strongly coupled sector along the lines of \cite{Intriligator:2006dd}.

Quite nicely, the most simple possible couplings of the Higgs and SUSY-breaking sectors in $F$-theory GUTs can realize precisely this scenario.  In particular, a $U(1)_{PQ}$ symmetry under which the field $X$ is charged naturally emerges from the geometry!  Moreover, as we shall see the effective action below the messenger scale is essentially of the "sweet spot" form \cite{Ibe:2007km}, meaning that we naturally land on a model which can be phenomenologically viable for suitable choices of parameters.

We also provide an example of how this scenario for gauge-mediated SUSY-breaking can be implemented in actual $F$-theory GUTs by using our approach to combine the Polonyi model of \cite{ourinst} with one of the $SU(5)$ GUT models of BHV II \cite{Beasley:2008kw}.
The result is a complete local model of an $SU(5)$ GUT with both MSSM matter and gauge-mediated supersymmetry breaking which realizes a simple mechanism for solving the $\mu$, $\mu/B_{\mu}$, and supersymmetric CP problems.  We also review what is needed to reproduce the successful phenomenology of \cite{Ibe:2007km}.  Though detailed numerics are not our aim, we find it amusing that that these conditions seem quite plausible.

During the course of this work we benefited from discussions with J.~Heckman and C.~Vafa, who were simultaneously interested in similar issues.  
We learned from them about the success of their F-theory construction \cite{HV} in providing a realization of the sweet spot supersymmetry breaking scenario.  This motivated us to reinvestigate our own earlier attempts at realizing it, leading to constructions that we understand to be very different from those of \cite{HV}.

The organization of this paper is as follows.  In section 2, we give a very brief review of some essential features of the BHV $F$-theory constructions \cite{Beasley:2008dc,Beasley:2008kw}.  In section 3, we discuss a basic framework for implementing gauge mediation in $F$-theory GUT models.  In section 4, we turn to the issue of coupling the SUSY-breaking and Higgs sectors and describe the natural way in which the $U(1)_{PQ}$ symmetry appears.  In section 5 we review the basic features of a simple D3-instanton triggered Polonyi model studied in \cite{ourinst}.  We then combine this with one of the $SU(5)$ GUT models of \cite{Beasley:2008kw} in section 6 to form a "complete" local model of gauge-mediated supersymmetry breaking that can address the $\mu$, $\mu/B_{\mu}$, and supersymmetric CP problems.  We comment on the ability of models of this type to realize the phenomenologically successful framework of \cite{Ibe:2007km} in section 7 before concluding in section 8.


\section{A brief review of F-theory GUT models}
\label{sec:FGUT}

\subsection{Bulk theory}
\label{subsec:FGUTbulk}

Here we give a very brief review of the essential ingredients used by BHV \cite{Beasley:2008dc,Beasley:2008kw} to build local GUT models in $F$-theory.
Start with F-theory \cite{Vafa:1996xn, Morrison:1996na, Morrison:1996pp} on an elliptically fibered Calabi-Yau four-fold $X$ with Calabi-Yau three-fold base $B$. Generically the elliptic fibration degenerates on a codimension one locus within $B$, which we denote by $S$ and in this section assume to be irreducible and compact.
When the degeneration along $S$ is locally of $A$ or $D$ type, such configurations can be described in IIB language as a collection of D7-branes wrapped on $S$ with possibly some O7 planes included as well \cite{Sen:1996vd}.  A novel feature of working directly in $F$-theory is the ability to describe $E$-type seven branes as well, making it possible to engineer gauge theories based on exceptional groups.
From the point of view of type IIB such compactifications are intrinsically non-perturbative.

In \cite{Beasley:2008kw}, it was argued that the spirit of bottom-up model building leads one to consider surfaces $S$ that are del Pezzo ($dP$).  The general philosophy is that one should study local models for which one could in principle take a strict decoupling limit $M_{Pl}\rightarrow\infty$, which separates GUT-scale physics from Planck-scale physics.  We shall adhere to this philosophy as well and hence will always assume that our surfaces are of $dP$ type.

The spectrum of the "bulk" theory on $S$ transforms in the adjoint of $G_S$. Switching on a gauge bundle $\mathcal{E}$ with structure group $H_S$
breaks the Lie algebra $\mathfrak{g}_S \rightarrow \mathfrak{h}_S \oplus \mathfrak{g}$, and thereby the adjoint representation into
\be
\mathfrak{g}_S = \bigoplus_i \rho_i \otimes  \sigma_i \,,
\ee
where $\rho_i$ ($\sigma_i$) are $\mathfrak{h}_S$ ($\mathfrak{g}$) representations.
The chiral spectrum transforming in a representation $\sigma_i$ of $\mathfrak{g}$ is determined by the bundle-valued Euler characteristic
\be
N_{\sigma_i} = -\chi_S (\mathcal{R}_i) \,,\qquad
N_{\sigma_i^*} = - \chi_S (\mathcal{R}_i^*) \,,
\ee
where $\mathcal{R}_i$ denotes the bundle transforming in $\rho_i$.
On a del Pezzo surface this is easily computed by
\be \label{dPcounting}
\chi_S(\mathcal{R}) =  1 - {1\over 2} c_1(\mathcal{R}) \cdot \mathcal{K}_S  + {1\over 2}
\Bigl(c_1^2(\mathcal{R})-2c_2(\mathcal{R})\Bigr) \,.
\ee
where $\mathcal{K}_S$ denotes the canonical class of $S$.
On a del Pezzo surface, various vanishing theorems preclude the existence of Yukawa couplings amongst bulk fields \cite{Beasley:2008dc} which requires that another source of  matter fields be introduced.


\subsection{Matter curves and brane-intersections}
\label{subsec:FGUTmatter}

Consider now two del Pezzo surfaces $S_1$ and $S_2$ intersecting along a complex curve $\Sigma$, so that the 7-branes wrapping the respective surfaces
intersect in a six-dimensional space.  Along $\Sigma$, the singularity type is enhanced to $G_{\Sigma}$ and, correspondingly, new bifundamental matter is localized there \cite{Bershadsky:1996nh,Katz:1996xe}.
To determine the specific matter content on the curve, we first decompose the adjoint of the enhanced $G_\Sigma$ gauge group with respect to the bulk gauge symmetries $G_{S_1}\times G_{S_2}$
\be\label{SigmaAdj}
\mathfrak{g}_\Sigma = \bigoplus_i (\rho^1_i  ,\rho^2_i) \,.
\ee
Representations other than the adjoints of $\mathfrak{g}_{S_1}$ and $\mathfrak{g}_{S_2}$ which appear in this decomposition determine the "bifundamentals" under which matter on $\Sigma$ will transform.  Each $G_{S_{1,2}}$ may then be broken by $U(1)$-bundles  $\mathcal{L}_{1,2}$ on $S_{1,2}$ to  $G_{S_{1,2}} \rightarrow U(1)_{1,2} \times G_{1,2}$, leading to a further decomposition
\be\label{SigmaDecomp}
(\rho^1, \rho^2) = \bigoplus_j (r^1_j, r^2_j)_{\alpha_j, \beta_j} \,,
\ee
where the $U(1)$ charges are denoted by $\alpha, \beta$ and $r^{1,2}$ are representations of $G_{1,2}$.

So far these were merely group-theoretic considerations for determining the representation content of the matter localized on $\Sigma$.  The actual matter spectrum, on the other hand, is determined by counting zero modes and this in turn is obtained by studying bundle-valued cohomologies.  In particular, the number $N_{(r_j^1,r_j^2)_{\alpha_j,\beta_j}}$ of zero modes in the representation $(r_j^1,r_j^2)_{\alpha_j,\beta_j}$ is given by \cite{Beasley:2008dc}
\be
N_{(r^1_j, r^2_j)_{\alpha_j, \beta_j}}
= h^0 (\Sigma, K_\Sigma^{1/2} \otimes \mathcal{L}_1^{\alpha_j}|_{\Sigma} \otimes {\mathcal{L}_2}^{\beta_j}|_{\Sigma}) \,,
\ee
where the restriction of bulk bundles to $\Sigma$ is denoted by $\mathcal{L}_{1,2}|_{\Sigma}$.  The net chirality on $\Sigma$ is also given by the simple relation \cite{Beasley:2008kw}
\be\label{NetChiral}
N_{(r^1_j, r^2_j)_{\alpha_j, \beta_j}}  -N_{\overline{(r^1_j, r^2_j)_{\alpha_j, \beta_j}}}
= deg \left( \mathcal{L}_1^{\alpha_j}|_{\Sigma} \otimes {\mathcal{L}_2}^{\beta_j}|_{\Sigma}\right) \,,
\ee
where $deg$ is the degree of the bundle.
These results can all be derived, for instance, by studying the six-dimensional defect theory living on the intersection of the 7-branes \cite{Beasley:2008dc}.


\subsection{Yukawa couplings}
\label{subsec:FGUTyukawa}

Of crucial importance for any model-building endeavor are the superpotential couplings between these various fields.  As discussed in \cite{Beasley:2008dc}, vanishing theorems on del Pezzo surfaces preclude the existence of superpotential couplings amongst bulk fields only.  Nontrivial couplings can arise, however, when matter curves $\Sigma_i$ intersect at isolated points where the singularity in the fiber is further enhanced.  This includes couplings between matter curve fields and bulk fields as well as couplings between matter curve fields only.  We shall focus on the latter type of coupling in this paper because none of our models will engineer charged matter in the bulk of any 7-branes.

At first glance, it might seem that Yukawa couplings amongst fields localized on matter curves are very hard to engineer.  This is because each such field is a bifundamental with respect to the gauge group of the bulk 7-branes on $S$ and the $U(1)$ on the additional 7-brane which intersects $S$ along $\Sigma$.  Even though the gauge boson on this additional 7-brane can easily be lifted{\footnote{Indeed, such $U(1)$'s are typically anomalous so are necessarily lifted by the Green-Schwarz mechanism.}}, the corresponding $U(1)$ still arises as a global symmetry of the action. As such, each matter field seems to come with its own independent $U(1)$ charge which must be respected in the superpotential.

However, in many cases not all of the $U(1)$'s on matter branes which meet at enhanced singular points are independent.  Rather various combinations are often identified, making nontrivial Yukawa couplings possible in cases where one might have naively thought otherwise.  Situations in which this happens typically do not have a simple perturbative description and hence must correspond to couplings that are generated nonperturbatively in type IIB.  Nevertheless, their presence is easy to see within F-theory from the direct analysis of \cite{Beasley:2008dc,Beasley:2008kw}.  Because this will play a crucial role throughout this paper, we now describe it in more detail in the context of a simple example.


\subsection{A Simple Example}

\subsubsection{Matter from $SU(2)$ Enhancement}

As an example. we consider now a single del Pezzo $S$ with an $I_1$ "singularity" corresponding, in the perturbative regime, to a single D7-brane.  We can engineer charged matter by enhancing the singularity to $SU(2)$ ($A_1$) along a curve $\Sigma$.  The geometry near $\Sigma$ can then be described by the unfolded $A_1$ singularity
\begin{equation}
y^2=x^2+z(z+t)\label{A1unfold} \,.
\end{equation}
As described in \cite{Beasley:2008dc}, the coordinates $x$, $y$, and $z$ of the fiber as well as the parameter $t$ are all sections of the canonical bundle $K_S$ over $S$.  For notational simplicity, though, we shall suppress any explicit dependence of these quantities on the coordinates of $S$.  Our original 7-brane sits at  $z=0$ and another now sits at $z+t=0$. They intersect along $\Sigma$, which lies along the locus $(z=0)\cap (z=-t)$.  Let us recall also that $t$ can be thought of as the expectation value of an $SU(2)$ adjoint field $\phi$ along $S$ whose nonzero value away from $\Sigma$ is responsible for breaking the gauge group $SU(2)\rightarrow U(1)_{\Sigma}$ \cite{Katz:1996xe,Beasley:2008dc}.  This breaking leads to bifundamental matter from the decomposition of the adjoint $\mathbf{3}$ of $SU(2)$ under 
\begin{equation}
\begin{aligned}
SU(2)       & \quad \rightarrow \quad  U(1)_{\Sigma} \cr
\mathbf{3}  & \quad \rightarrow \quad  \mathbf{1}_0\oplus\mathbf{1}_{+2}\oplus\mathbf{1}_{-2} \,.
\label{su2adjd}
\end{aligned}
\end{equation}
The factor $\mathbf{1}_0$ above simply reflects the adjoint of $U(1)_{\Sigma}$ so we identify $\mathbf{1}_{+2}$ as the bifundamental representation that is engineered.  The matter in this representation is localized along that part of $z=0$ where $t=0$.  In other words, it is localized on $\Sigma$.

We can visualize this configuration also in terms of type IIB objects as, in the perturbative limit, it reduces to a pair of D7 branes which intersect along $\Sigma$.  Locally, one can obtain such a configuration by starting with parallel D7 branes and then rotating one of them.  This rotation can be achieved by giving a varying expectation value to the adjoint scalar field which increases as one moves away from $\Sigma$.  This is the perturbative analog of deforming the geometry \eqref{A1unfold} by letting $t$ be nonzero away from $\Sigma$ on $S$.  After this rotation, the total gauge group on the D7 branes is $U(1)_S\times U(1)_a$ and the bifundamentals carry charge $(+,-)$.  Comparing with \eqref{su2adjd}, we see that $U(1)_{\Sigma}$ should be identified with the specific linear combination of $U(1)_S\times U(1)_a$ with respect to which the bifundamentals are charged
\begin{equation}
Q_{\Sigma}=Q_S - Q_a\,.
\end{equation}
The overall diagonal $U(1)$, with respect to which the bifundamental matter is uncharged, is absent from the F-theory description{\footnote{In the $SU(5)$ GUTs of \cite{Beasley:2008dc,Beasley:2008kw}, it is this lack of overall $U(1)$ which allows one to engineer the $\mathbf{10}\times\mathbf{10}\times\mathbf{5}$ couplings that are perturbatively forbidden \cite{Blumenhagen:2001te} in intersecting brane models.}}.

\subsubsection{Yukawa Couplings from an $SU(3)$ Point}
\label{subsubsec:yukawasu3}

We turn now to isolated singularities where matter curves can meet.  Distinct $SU(2)$ curves, for instance, can intersect at points where the singularity is further enhanced by one rank to $SU(3)$.  The local geometry near such a point takes the form
\begin{equation}
y^2=x^2+z(z+t_1)(z+t_2) \,,
\label{A2sing}
\end{equation}
with $t_1=t_2=0$ defining the $SU(3)$ enhanced point.  This corresponds to three D7-branes, namely our original one at $z=0$ and two additional "matter branes" along $z+t_1=0$, and $z+t_2=0$.  Note that there are generically three curves of $SU(2)$ enhancement, namely $t_1=0$, $t_2=0$, and $t_3\equiv t_1-t_2=0$.  The first two correspond to curves where the "matter branes" intersect the $z=0$ 7-brane and we denote them by $\Sigma_1$ and $\Sigma_2$, respectively.  The third, $t_3=0$, is simply the intersection of the "matter branes" with one another and is denoted by $\Sigma_3$ in what follows.  Following any given matter curve toward the $SU(3)$ singularity specifies an embedding of its gauge group, $U(1)_i$, into $SU(2)_i$ and then further into $SU(3)$.  Because $SU(3)$ has rank 2, there are only two independent such embeddings.  This means that the $U(1)_i$'s under which matter on the $\Sigma_i$ is charged must satisfy a nontrivial relation.  This is captured by the fact that the deformation parameters $t_i$ are not independent but instead satisfy  $t_3=t_1-t_2$.  In fact, if we recall that the $t_i$ correspond to elements of the Cartan subalgebra of $SU(3)$ which are in turn identified with expectation values of an $SU(3)$ adjoint field $\phi$ \cite{Katz:1996xe,Beasley:2008dc}, it is not hard to see that $U(1)_i$ is simply the $U(1)$ subgroup of $SU(3)$ that is generated by $t_i$.

Given this, we turn now to the charges of various fields with respect to a fixed choice of two independent $U(1)$'s, which we take to be $U(1)_1$ and $U(1)_2$.  Following \eqref{su2adjd}, we see that fields localized on $\Sigma_1$ have charge $(\pm 2,0)$ under $U(1)_1\times U(1)_2$ while fields localized on $\Sigma_2$ carry instead charge $(0,\pm 2)$.  Fields localized on $\Sigma_3$ have charges $\pm 2$ with respect to $U(1)_3$ but, as we saw before, the generator of this $U(1)$ is not independent of $t_1$ and $t_2$ but rather is given simply by the difference $t_1-t_2$.  As such, fields on $\Sigma_3$ carry charges $(2,-2)$ and $(-2,2)$ under $U(1)_1\times U(1)_2$.  This means that nonzero Yukawa couplings which are invariant under both $U(1)_1$ and $U(1)_2$ can be obtained by combining fields from all three of the matter curves that meet at the $SU(3)$ point.

Note that we could see this directly by simply decomposing the adjoint $\mathbf{8}$ of $SU(3)$
\begin{equation}
\begin{aligned}
SU(3)   \quad &\rightarrow \quad U(1)_1\times U(1)_2 \cr
\mathbf{8}\quad &\rightarrow \quad \mathbf{1}_{0,0}\oplus\mathbf{1}_{0,0}\oplus\left(\mathbf{1}_{2,0}\oplus\mathbf{1}_{-2,0}\right)\oplus\left(\mathbf{1}_{0,2}\oplus\mathbf{1}_{0,-2}\right)\oplus\left(\mathbf{1}_{-2,2}\oplus\mathbf{1}_{2,-2}\right) \,.
\end{aligned}
\end{equation}
We identify the two factors of $\mathbf{1}_{0,0}$ as the adjoint of $U(1)_1\times U(1)_2$.  Each quantity in parentheses then represents bifundamental matter associated to a matter curve that can emanate from the $SU(3)$ enhancement point.  There are three such collections and hence an $SU(3)$ point generically describes the intersection of three matter curves.  This decomposition also gives the $U(1)$ charges for all three sets of fields expressed in a single basis so that it is clear what gauge invariant Yukawa couplings can originate at the $SU(3)$ point.  In this case, we can have either $\mathbf{1}_{2,0}\otimes\mathbf{1}_{0,-2}\otimes\mathbf{1}_{-2,2}$ or its conjugate.

Note that such couplings are precisely what we expect from triple intersections of D7-branes in the perturbative type IIB description.  In particular, the matter fields $\mathbf{1}_{2,0}$ and $\mathbf{1}_{0,-2}$ simply correspond to bifundamentals connecting the $z=0$ brane to the "matter branes" while $\mathbf{1}_{-2,2}$ is the bifundamental which connects the "matter branes" to one another.  That $\mathbf{1}_{-2,2}$ is a singlet under the $U(1)_3$ gauge group on the $z=0$ brane follows from its relation to $U(1)_1$ and $U(1)_2$, namely $t_3=t_1-t_2$.

While we might have expected the presence of three D7-branes to lead to 3 independent $U(1)$'s which restrict the form of the Yukawa couplings, we see that only two make an appearance in the F-theory description.  In this simple example, the $U(1)$ that is not present is the overall diagonal $U(1)$ with respect to which none of the bifundamental fields carry a net charge.  Its absence is easily understood because this $U(1)$ is expected to decouple even from the perturbative point of view.  As described in \cite{Beasley:2008dc}, however, this nontrivial identification of $U(1)$'s persists also for $D$ and $E$ type enhancements where the interpretation is not as trivial.  As such one finds allowed couplings which, in the case of $E$-type enhancements, are perturbatively forbidden in type IIB{\footnote{The most notable example of this is the $\mathbf{5}\times\mathbf{10}\times\mathbf{10}$ Yukawa coupling of $SU(5)$ GUTs which is perturbatively forbidden in intersecting brane models \cite{Blumenhagen:2001te} but can be generated there by instanton effects \cite{Blumenhagen:2007zk}.}}.

This simple example serves to demonstrate the well-known connection between group theory and geometry in this class of local Calabi-Yau which allows the above procedure for determining Yukawa couplings to be applied quite generally.  Given an isolated point with singularity $G$, a simple decomposition of the adjoint indicates both the kind of matter curves which can meet there and the nature of the Yukawa couplings that can be generated.  We shall make extensive use of this fact in all that follows.


\section{The Messenger Sector}
\label{sec:GM}

\begin{figure}\begin{center}
\subfigure[GUT Brane with $f$ and $\bar{f}$ matter curves $\qquad$
and the singlet $X$]
{\epsfig{file=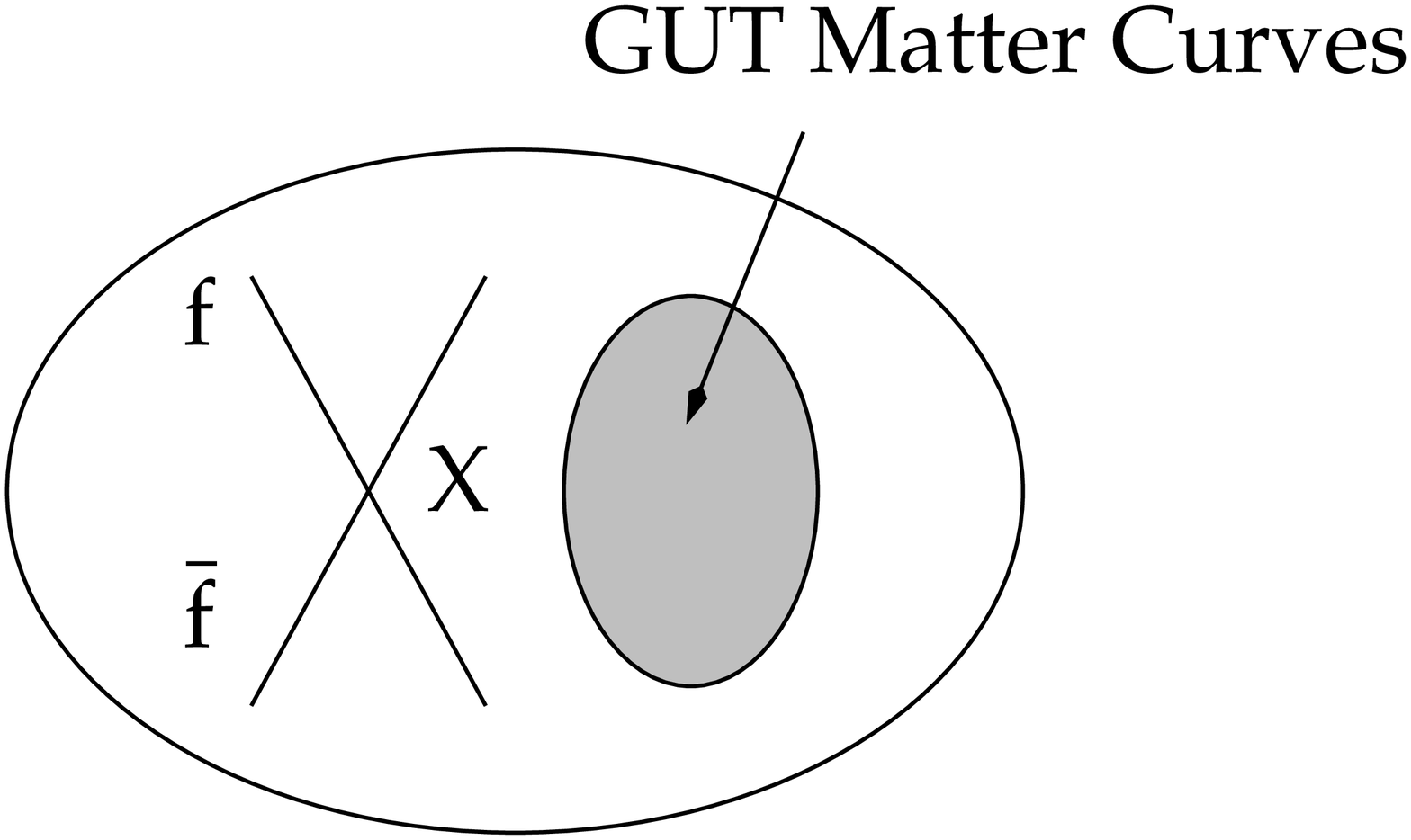,width=.45\textwidth
}\label{ffbarsetup}}
\subfigure[Depiction of the triple intersection at the $SU(7)$ $\qquad $ enhancement point]
{\epsfig{file=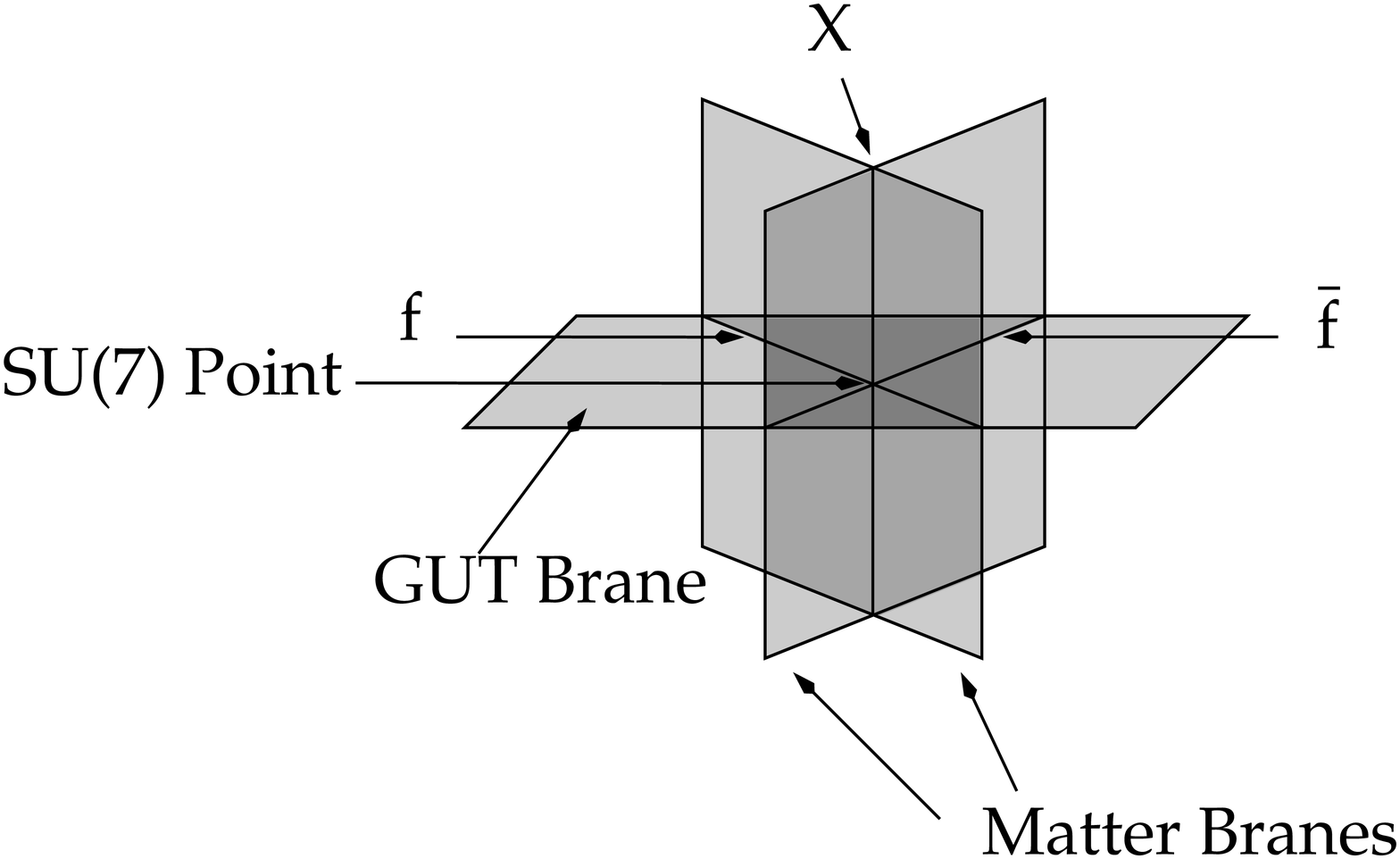,width=0.53\textwidth
}\label{intbranes}}
\label{GUTcouplingsetup}
\caption{Basic setup for coupling Polonyi to $F$-theory GUTs}
\end{center}
\end{figure}

In this section, we discuss a simple way to incorporate gauge-mediated supersymmetry breaking into F-theory GUTs.  While the basic idea of this approach has already appeared in BHV II \cite{Beasley:2008kw}, we review it here and emphasize that it naturally incorporates one of the D3-instanton triggered SUSY-breaking models of \cite{ourinst}.  We shall defer a detailed review of this model to section \ref{sec:SusyBreaking} and its incorporation in a sample gauge-mediated model to section \ref{subsec:PolonyiGUT}.

Let us suppose that we want to communicate SUSY-breaking to an F-theory GUT model with charged messenger fields, $f$ and $\bar{f}$, transforming in the $\mathbf{5}$ and $\mathbf{\overline{5}}$, respectively, of $SU(5)$.  One way to introduce such fields is to add a new pair of matter curves, $\Sigma_f$ and $\Sigma_{\bar{f}}$, to the GUT brane.  These curves correspond to local $SU(6)$ enhancements of the $SU(5)$ singularity on the GUT brane.  To obtain a nontrivial interaction between $f$ and $\bar{f}$ these two matter curves must intersect at an isolated $SU(7)$ singularity.

This setup, depicted in figure \ref{intbranes}, is now very similar to the mechanism proposed in \cite{Beasley:2008kw} for generating a $\mu$ term.  In particular, the $SU(7)$ singularity describes the standard triple intersection of three D7-brane stacks that we are accustomed to in the perturbative type IIB language.  The fields $f$ and $\bar{f}$ are bifundamentals connecting the matter branes to the GUT brane.  In addition, however, we get one more field which is a bifundamental connecting the matter branes to one another.  One can also see this more directly from the decomposition of the $SU(7)$ adjoint under
\begin{equation}
\begin{aligned}
SU(7)   \quad & \rightarrow  \quad SU(5)\times U(1)\times U(1)\cr
\mathbf{48}\quad &\rightarrow  \quad \left(\mathbf{24}_{0,0}\oplus\mathbf{1}_{0,0}\oplus\mathbf{1}_{0,0}\right)\oplus\left(\mathbf{5}_{0,6}\oplus\mathbf{\overline{5}}_{0,-6}\right)\oplus\left(\mathbf{5}_{-7,1}\oplus\mathbf{\overline{5}}_{7,-1}\right)\oplus\left(\mathbf{1}_{7,5}\oplus\mathbf{1}_{-7,-5}\right) \,,
\end{aligned}
\end{equation}
where we use the $U(1)$ charge conventions of Slansky \cite{Slansky:1981yr}.  The chiral multiplet $X$ is a GUT singlet as its matter curve intersects the GUT brane only at the $SU(7)$ point.  The interaction that we obtain from this point is quite familiar as it is the standard one from ordinary gauge mediation
\begin{equation}
W_{OGM}\sim Xf\bar{f}\label{WOGM} \,,
\end{equation}
provided $X$ picks up a SUSY-breaking expectation value in its $F$-component.  In the spirit of \cite{Beasley:2008dc,Beasley:2008kw}, we could now simply assume that some physics associated with the $f$ and $\bar{f}$ branes imposes this condition and thereby take it as input in our F-theory GUT.

Quite remarkaby, however, this inclusion of messengers and spurion field $X$ is identical to what was needed to engineer a very simple Polonyi model of supersymmetry-breaking in \cite{ourinst}.  In particular, it was shown that with suitable choices of flux on the $f$ and $\bar{f}$ matter branes, D3-instantons will automatially trigger SUSY-breaking at an exponentially small scale!

We will review the construction of the Polonyi model of \cite{ourinst} later in section \ref{sec:SusyBreaking} and discuss its coupling to F-theory GUTs in more detail when building a "complete" model in section \ref{sec:completemodel}.  For the general discussion of gauge mediation that follows, however, we will simply presume that some dynamics on the matter branes cause the field $X$ to pick up both scalar and $F$-component expectation values
\begin{equation}
\langle X\rangle  = M + \theta^2 F_X \,,
\end{equation}
which, through the coupling \eqref{WOGM}, gives a mass to the messengers $f$ and $\bar{f}$ and breaks supersymmetry.


\section{Higgs Sector and Generation of $\mu$}
\label{sec:HiggsMu}

Any model of gauge mediated supersymmetry breaking must address the $\mu$ and $\mu/B_{\mu}$ problems.  In the first part of this section, we will briefly review these issues as well as an elegant solution due to Ibe and Kitano which utilizes $U(1)_{PQ}$ symmetry \cite{Ibe:2007km}.  We will then demonstrate that this solution arises completely naturally when gauge mediation is incorporated into F-theory GUT models.


\subsection{The $\mu$ and $\mu/B_{\mu}$ Problems and $U(1)_{PQ}$}
\label{subsec:muBmu}

A crucial issue faced by any model in which gauge mediation dominates is an explanation for why $\mu$ sits naturally near the electroweak scale rather than at the Planck scale.  A common approach to this issue is to assume that the $\mu$ parameter vanishes at high scales and is generated at low scales by the same physics that breaks supersymmetry.  This can be implemented, for example, by coupling the Higgs directly to the messenger fields in the superpotential or some other suitably heavy fields which also couple to $X$.  Integrating out these massive fields then generates the effective operators
\begin{equation}
\frac{1}{M}\int\,d^4\theta\,H\bar{H}X^{\dag}\qquad\text{and}\qquad \frac{1}{M^2}\int\,d^4\theta\,H\bar{H}XX^{\dag}\label{muBmuops} \,.
\end{equation}
When the $F$-component of $X$ picks up a nonzero expectation value, the first of these gives rise to a $\mu$ term and the second to a $B_{\mu}$ term.

Naturalness of electroweak symmetry breaking requires $\mu$ and $B_{\mu}$ to be at roughly the same scale
\begin{equation}
B_{\mu}\sim \mu^2\,.
\end{equation}
Given \eqref{muBmuops} it seems as though this is easily achieved.  However, in \eqref{muBmuops} we have neglected to write the loop suppression factor $1/16\pi^2$ that arises when generating these operators by integrating out heavy fields.  In general, both operators are generated at the same loop order so both $\mu$ and $B_{\mu}$ pick up one factor of $1/16\pi^2$.  This means that $B_{\mu}$ is in fact larger than $\mu^2$ by a factor of about $10^2$, introducing an extra fine-tuning that has been dubbed the $\mu/B_{\mu}$ problem \cite{Dvali:1996cu}.

One nice way to address the $\mu$ problem is to introduce a $U(1)_{PQ}$ symmetry under which $H$ and $\bar{H}$ both have charge +1.  Such a symmetry forbids the appearance of a bare $\mu$ term in the superpotential and is often invoked in an approximate form to explain why $\mu$ is naturally small.  If we also suppose that $X$ carries nonzero $PQ$ charge \cite{Ibe:2007km}, then the $F$-component expectation value which breaks supersymmetry will also break $U(1)_{PQ}$ at the same scale.  In fact, the first operator of \eqref{muBmuops} becomes allowed provided we specify the $PQ$ charge of $X$ to be +2.  In this scenario, $\mu$ is thus naturally generated with the same exponential suppression factor that arises in the breaking of supersymmetry.
Furthermore, the $U(1)_{PQ}$ symmetry continues to forbid the second operator of \eqref{muBmuops} so that $B_{\mu}=0$ at the messenger scale.  This is a highly predictive scenario which has received a great deal of attention in the literature \cite{Babu:1996jf,Dimopoulos:1996yq,Bagger:1996ei,Rattazzi:1996fb,Borzumati:1996qs,Gabrielli:1997jp}.  For our purposes, it suffices to note that, in this case, a $B_{\mu}$ parameter of the right size is generated by MSSM RG running below the messenger scale (see, for instance, \cite{Babu:1996jf}).  Quite happily, the CP phase $\text{arg}(m_{1/2}\mu(B_{\mu})^{\ast})$ also vanishes at the messenger scale in this scenario, leading to a successful resolution to the supersymmetric CP problem \cite{Ibe:2007km}.

In what follows, we shall mainly be interested in demonstrating that the $U(1)_{PQ}$ symmetry that played such a crucial role in the above story arises naturally when gauge mediation is incorporated into F-theory GUT models.  We shall also make some comments about numerics and the ability to reproduce the specific framework of \cite{Ibe:2007km} in section \ref{sec:numerology} but we will not make any sharp statements about the values of $\mu$ or any other soft parameters because they will depend on dimensionless coefficients that we cannot compute.


\subsection{SUSY-Breaking and the Higgs Sector in F-theory GUTs}
\label{subsec:SBHiggs}

We now turn to the generation of $\mu$ in the gauge mediation framework of section \ref{sec:GM}.  As discussed above, this necessitates a direct coupling between the Higgs and messenger sectors.  Because approximate $U(1)$ symmetries are quite plentiful in the BHV formalism it seems reasonable to expect that a $U(1)_{PQ}$ symmetry of the sort described above can be obtained in this context.  One potential pitfall, however, is that each multiplet typically has its own matter brane and hence its own $U(1)$ charge.  What saves us is that, as mentioned in section \ref{subsubsec:yukawasu3}, not all of these $U(1)$'s remain independent when matter curves participate in triple intersections at enhanced singular points.
Because of this, a $U(1)_{PQ}$ under which all of $H$, $\bar{H}$, and $X$ are charged can in principle arise.  In fact, we will see that such a symmetry arises completely naturally.

The simplest way to engineer a coupling between the Higgs and messenger sectors is to require the  Higgs and messenger matter curves to intersect one another.  Because the Higgs and messenger fields all transform in the $\mathbf{5}$ or $\mathbf{\overline{5}}$ of $SU(5)$, these matter curves all correspond to local $SU(6)$ enhancements.  It is easy to see that two such curves can intersect at isolated points where the singularity enhances to either $SO(12)$ or $SU(7)${\footnote{The other possible rank one enhancement is to $E_6$.  However, such points describe the intersection of 2 $\mathbf{10}$ matter curves with one $\mathbf{5}$ matter curve so are not relevant here.}}.  We will now consider each of these possibilities in turn.


\subsection{$SO(12)$ Enhancement}
\label{subsec:SO12}

We first consider the possibility that the Higgs and messenger curves meet at points of $SO(12)$ enhancement.  To see what type of couplings can be generated there, consider the decomposition of the $SO(12)$ adjoint under
\begin{equation}
\begin{aligned}\label{SO12adj}
SO(12) & \quad \rightarrow\quad  SU(5)\times U(1)_1\times U(1)_2 \cr
\mathbf{66} & \quad \rightarrow\left(\mathbf{24}_{0,0}\oplus\mathbf{1}_{0,0}\oplus\mathbf{1}_{0,0}\right)\oplus\left(\mathbf{5}_{2,2}\oplus\mathbf{\overline{5}}_{-2,-2}\right)\oplus \left(\mathbf{5}_{-2,2}\oplus\mathbf{\overline{5}}_{2,-2}\right)\oplus\left(\mathbf{10}_{0,4}\oplus\mathbf{\overline{10}}_{0,-4}\right) \,.
\end{aligned}
\end{equation}
We see from this that isolated $SO(12)$ singularities generically occur at the intersection of two $\mathbf{5}$ matter curves and a $\mathbf{10}$ matter curve.  Couplings that originate at such a triple intersection must respect the $U(1)_1\times U(1)_2$ symmetry and hence take the form $\mathbf{5}\times \mathbf{5}\times\mathbf{\overline{10}}$ or its conjugate.  This means that if we want a nontrivial interaction between two fields localized on $\mathbf{5}$ matter curves which meet at an $SO(12)$ point, it is necessary to introduce an additional $\mathbf{10}$ matter curve.

Of course, in the minimal setup where each of our two messenger matter curves meets one of the Higgs matter curves, we will have two singular points.  If both are $SO(12)$ enhancements then the simplest possibility which yields a nontrivial interaction at each is to have a single $\mathbf{10}$ matter curve connecting the two{\footnote{As shall become more clear later, having a single $\mathbf{10}$ matter curve intersecting both is actually crucial for generating $\mu$. If we introduced two separate $\mathbf{10}$ matter curves there would be an extra $U(1)$ which in fact prevents $\mu$ from being generated at all.}}.

\begin{figure}
\begin{center}
\epsfig{file=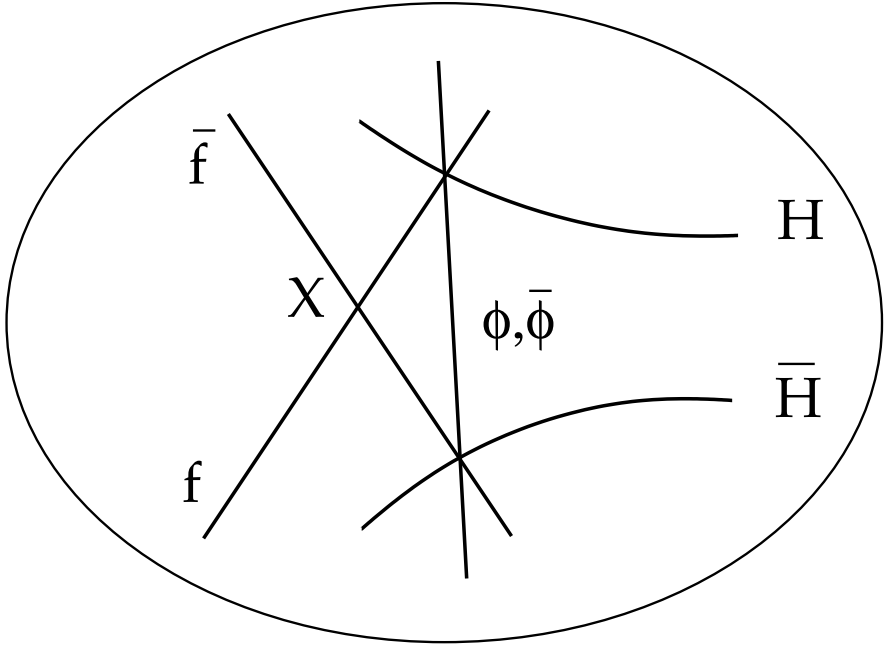,width=0.7\textwidth}
\caption{Model of gauge mediation with coupling to Higgs sector at points of $SO(12)$ enhancement that leads to small $\mu$ and $B_{\mu}=0$.}
\label{gettingmu}
\end{center}
\end{figure}

These considerations motivate us to consider the general setup of figure \ref{gettingmu}.  Because we ultimately want to integrate out the extra fields on the $\mathbf{10}$ matter curve, it is important that they become massive.  There is a simple mechanism for this at our disposal, though, namely to choose a bundle on the matter brane that  eliminates all $\mathbf{10}$ and $\mathbf{\overline{10}}$ zero modes.  In that case, the lightest fields localized there are KK modes with GUT scale masses and the only nontrivial tree level interaction among zero modes in figure \ref{gettingmu} is simply that of ordinary gauge mediation \eqref{WOGM}.  The SUSY-breaking sector is indeed directly coupled to the Higgs sector but only by physics at the GUT scale.

The effect of integrating out the various KK modes in this scenario is to generate higher dimension operators in the effective theory for the spurion field, $X$, the messengers, $f$ and $\bar{f}$, and the Higgs fields, $H$ and $\bar{H}$.  As usual, the specific operators that can be generated are determined by the relevant set of global symmetries.  To determine these, we turn our attention to the form of the full tree-level superpotential that arises from figure \ref{gettingmu} including couplings involving KK modes.  After that, we will explicitly show how these symmetries arise from the geometry.

Using $\phi,\bar{\phi}$ to denote KK modes on the $\mathbf{10}$ matter curve, we can write the superpotential associated to figure \ref{gettingmu} as
\begin{equation}W\sim Xf\bar{f}+Hf\bar{\phi}+\bar{H}\bar{f}\phi+M_{GUT}\phi\bar{\phi}+\ldots\label{so12sup}\end{equation}
Included in the $\ldots$ are couplings similar to the above but with some or all of $H,\bar{H},f,\bar{f}$ replaced by KK modes on the corresponding matter curve.  To see what kind of terms can be generated by integrating out the KK modes we note that, quite nicely, the superpotential \eqref{so12sup} is invariant under a $U(1)_{PQ}$ symmetry under which the various fields carry charges
\begin{equation}
\begin{array}{c|c|c|c|c|c|c|c}
 & X & f & \bar{f} & \phi & \bar{\phi} & H & \bar{H} \\ \hline
 U(1)_{PQ} & 2 & -1 & -1 & 0 & 0 & 1 & 1\\
 \end{array}
\label{U1PQ} \end{equation}
This is precisely what we needed for the mechanism of section \ref{subsec:muBmu} to work!  Indeed, we see that the operator
 \begin{equation}
 \frac{1}{M_{GUT}}\int\,d^4\theta\,X^{\dag}H\bar{H}\,,
 \label{muop}\end{equation}
 is allowed and leads to the generation of a $\mu$-term 
\begin{equation}
 \mu\sim \frac{F_X}{M_{GUT}} \,,
 \end{equation}
Moreover, it is easy to verify directly from the form of \eqref{so12sup} that loops of KK modes can generate the operator \eqref{muop}.  Because there are numerous modes in the KK tower with a variety of different Yukawa couplings,
though, we are not currently able to reliably compute the coefficient which appears here.  What is important for our purposes, instead, is the appearance of $F_X$ which makes manifest that the instanton-generated scale enters, leading to the desired exponential suppression.

In fact, we can go one step further and write down all of the operators which are generated up to and including dimension 6
\begin{equation}
\ba
\delta {\cal{L}}
\sim &\int\,d^4\theta\left(\frac{1}{M_{GUT}}X^{\dag}H\bar{H}+\frac{1}{M_{GUT}^2}X^{\dag}X(H H^\dag + \bar{H}\bar{H}^\dag)+\ldots\right) \cr
&\qquad \qquad +\int\,d^2\theta\left(\frac{1}{M_{GUT}}H\bar{H}f\bar{f}+\ldots\right) \,.
\ea
\label{highdimops}\end{equation}
Further integrating out the messengers, $f$ and $\bar{f}$, simply gives an additional contribution to the coefficient of the operator \eqref{muop} which is proportional to $\ln (M_{GUT} / M_{\text{Mess}})$.  As such, we land on an effective action of precisely the same form as that of Ibe and Kitano's "sweet spot supersymmetry" \cite{Ibe:2007km}.  
Among the benefits of this model is the fact 
the operator
 \begin{equation}
 \int\,d^4\theta\,X^{\dag}X H\bar{H}
 \end{equation}
 is forbidden so that $B_{\mu}$ is not generated.  As discussed in \cite{Ibe:2007km}, this can provide a natural solution to both the $\mu/B_{\mu}$ and supersymmetric CP problems.

Before we move on, it is important to note that figure \ref{gettingmu} represents in fact one of two possible choices we could have made to couple the messenger and Higgs sectors at a pair of $SO(12)$ enhancements with only one extra matter curve.  Alternatively, we could have interchanged the $f$ and $\bar{f}$ matter curves.  In this case, gauge invariance would preclude any direct couplings between the messenger fields $f,\bar{f}$ and the Higgs fields, $H,\bar{H}$.  Though the Higgs fields are still coupled to the SUSY-breaking field $X$ through loops of KK modes, it is not difficult to see that the $PQ$ charge of $X$ in this case is flipped so that the operator $\int\,d^4\theta\,X^{\dag}H\bar{H}$ is forbidden and hence $\mu$ is not generated.


\subsubsection{$U(1)_{PQ}$ From Geometry}
\label{subsubsec:SO12PQ}

As we have repeatedly emphasized, the $U(1)_{PQ}$ symmetry of \eqref{so12sup} plays a crucial role in connecting the generation of $\mu$ to SUSY-breaking while simultaneously forbidding the generation of $B_{\mu}$.  Typically, imposing $U(1)$ symmetries such as this fixes the form of the superpotential that one writes down.  In these $F$-theory constructions, however, it is the geometry which unequivocally determines the form of the superpotential \eqref{so12sup}.  As such, it must be possible to see directly how the $U(1)$ symmetries which constrain the form of the superpotential can arise from the geometry.  In this subsection, we demonstrate this simple idea for the gauge mediated model of figure \ref{gettingmu} in order to see the emergence of $U(1)_{PQ}$.

We start by recalling that each matter brane which engineers a field $\Phi$ has its own gauge group, $U(1)_{\Phi}$.  In the conventions of Slansky \cite{Slansky:1981yr}, the charges of various $SU(5)$ fields that we can engineer under their corresponding matter branes are given by

\begin{center}\begin{tabular}{c|ccc}
Field & $U(1)_{H}$  & $U(1)_{f}$ & $U(1)_{\bar{\phi}}$ \\
\hline
$H$ & $6$& $0$& $0$  \\
$f$ &  $0$ & $6$ & $0$\\
$\bar{\phi}$ & $0$ & $0$ & $-4$
\end{tabular}
\end{center}

As we see from the decomposition \eqref{SO12adj}, when three matter branes meet at an $SO(12)$ point there are only two independent $U(1)$'s under which the fields are charged.
In particular, we read off from \eqref{SO12adj} that the three bifundamentals which can interact at such a singularity are either
\begin{equation}
\mathbf{5}_{2,2}\times\mathbf{5}_{-2,2}\times\mathbf{\overline{10}}_{0,-4} \,,
\end{equation}
or the conjugates, where we have listed the $U(1)_1\times U(1)_2$ charges.  Let us denote these fields by
\begin{equation}
H\sim \mathbf{5}_{2,2}\qquad f\sim \mathbf{5}_{-2,2}\qquad \bar{\phi}\sim\mathbf{\overline{10}}_{0,-4} \,.
\end{equation}
The charges under $U(1)_1$ and $U(1)_2$ are now easily identified as the following combinations of the matter brane $U(1)$'s
\begin{equation}
Q_1 = \frac{1}{3}\left(Q_H - Q_f\right)\qquad Q_2=\frac{1}{3}\left(Q_H+Q_f\right)+Q_{\bar{\phi}}\,.
\end{equation}

We see something similar at the $\bar{H}\bar{f}\phi$ intersection point.  There, if we denote the two $U(1)$'s at the $SO(12)$ singularity by $U(1)_3$ and $U(1)_4$ we find that
\begin{equation}
Q_3 = \frac{1}{3}\left(Q_{\bar{H}}-Q_{\bar{f}}\right)\qquad Q_4 = \frac{1}{3}\left(Q_{\bar{H}}+Q_{\bar{f}}\right)+Q_{\phi}\,.
\end{equation}
Because we started with 5 $U(1)$'s, namely $U(1)_H$, $U(1)_{\bar{H}}$, $U(1)_f$, $U(1)_{\bar{f}}$, and $U(1)_{\phi}$ the loss of a $U(1)$ at each $SO(12)$ singularity should leave us with only three.
It appears at the moment that we have four but this is because we have not properly "glued" the two singularities together by identifying $U(1)_{\bar{\phi}}$ with $U(1)_{\phi}$ (with the appropriate sign of course).  This is also easily done and leaves us with three $U(1)$'s corresponding to $U(1)_1$, $U(1)_3$, and a third $U(1)$, which we refer to as $U(1)_C$
\begin{equation}
Q_C = \frac{1}{3}\left(Q_H+Q_{\bar{H}}+Q_f+Q_{\bar{f}}\right)-Q_{\phi} \,.
\end{equation}
Defining also
\begin{equation}
Q_A=\frac{Q_1-Q_3}{2}\qquad\text{and}\qquad Q_B=\frac{Q_1+Q_3}{2}\,,
\end{equation}
we can now list the charges of our fields under a choice of 3 independent $U(1)$'s as

\begin{center}\begin{tabular}{c|c|c|c}
Field & $U(1)_A$ & $U(1)_B$ & $U(1)_C$ \\ \hline
$H$ & 1 & 1 & 2 \\
$\bar{H}$ & 1 & -1 & -2 \\
$f$ & -1 & -1 & 2\\
$\bar{f}$ & -1 & 1 & -2 \\
$\phi$ & 0 & 0 & -4 \\
$\bar{\phi}$ & 0 & 0 & 4 \\
$X$ & 2 & 0 & 0
\end{tabular}\end{center}

We immediately recognize $U(1)_A$ as our Peccei-Quinn symmetry, $U(1)_{PQ}$, from \eqref{U1PQ}.  Moreover, it is easy to verify that the superpotential \eqref{so12sup} is indeed the most general one that can be written down which preserves the full $U(1)_A\times U(1)_B\times U(1)_C$ symmetry.  Although it was expected at the outset, we find it gratifying to see, in the context of a simple example, the connection between geometry and global symmetries of the effective action.


\subsection{$SU(7)$ Enhancement}
\label{subsec:SU7}

\begin{figure}
\begin{center}
\epsfig{file=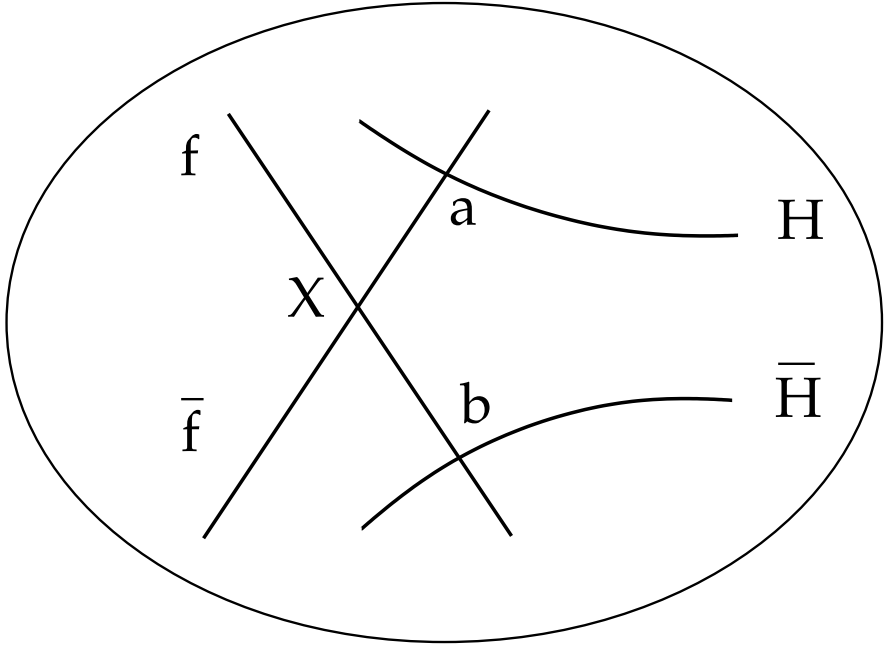,width=0.7\textwidth}
\caption{Model of gauge mediation with coupling to Higgs sector at points of $SU(7)$ enhancement that leads to small $\mu$ and $B_{\mu}=0$ provided $a$ and $b$ pick up nonzero expectation values.}
\label{su7mu}
\end{center}
\end{figure}

An alternative choice for the Higgs and messenger curve intersections is a local $SU(7)$ enhancement.  We have already discussed the properties of these points in detail when constructing the messenger sector in section \ref{sec:GM}.  As we saw there, the decomposition of the $SU(7)$ adjoint under
\begin{equation}\begin{aligned}
SU(7) &\quad \rightarrow \quad SU(5)\times U(1)\times U(1) \cr
\mathbf{48}& \quad \rightarrow\quad  \left(\mathbf{24}_{0,0}\oplus\mathbf{1}_{0,0}\oplus\mathbf{1}_{0,0}\right)\oplus\left(\mathbf{5}_{0,6}\oplus\mathbf{\overline{5}}_{0,-6}\right)\oplus\left(\mathbf{5}_{-7,1}\oplus\mathbf{\overline{5}}_{7,-1}\right)\oplus\left(\mathbf{1}_{7,5}\oplus\mathbf{1}_{-7,-5}\right) \,,
\end{aligned}
\end{equation}
suggests that we can get nonzero interactions of the form $\mathbf{5}\times \mathbf{\overline{5}}\times\mathbf{1}$ with the $\mathbf{1}$ being a GUT singlet which is a bifundamental with respect to the two matter branes.  Because we do not need to introduce any new matter curves on the GUT brane to get nontrivial interactions at the $SU(7)$ points, we thus consider the minimal setup in figure \ref{su7mu}{\footnote{A setup with one $SO(12)$ point and one $SU(7)$ point is also straightforward but contains no essential new ingredients.}}.

It is now a simple matter to write the superpotential associated to figure \ref{su7mu}.  Denoting the new singlet fields at the $SU(7)$ intersections by $a$ and $b$ we have
\begin{equation}
W\sim Xf\bar{f} + aH\bar{f} + b\bar{H} f\label{su7sup}\,.
\end{equation}
Once again, this superpotential is invariant under a $U(1)_{PQ}$ symmetry with charges
\begin{equation}\begin{array}{c|c|c|c|c|c|c|c} & X & f & \bar{f} & a & b & H & \bar{H} \\ \hline
U(1)_{PQ} & 2 & -1 & -1 & 0 & 0 & 1 & 1\end{array}\label{U1PQ2}\end{equation}
Unfortunately, it possesses other $U(1)$ symmetries which will give us some trouble.  In particular, the superpotential \eqref{su7sup} is specified by $U(1)_{PQ}$ and three additional symmetries, $U(1)_a$, $U(1)_b$, and $U(1)_c$ with charges
\begin{equation}\begin{array}{c|c|c|c|c|c|c|c}
 & X & f & \bar{f} & a & b & H & \bar{H} \\ \hline
U(1)_a & 0 & 0 & 0 & 1 & 0 & -1 & 0 \\
U(1)_b & 0 & 0 & 0 & 0 & 1 & 0 & -1 \\
U(1)_c & 0 & 1 & -1 & 0 & 0 & 1 & -1
\end{array}\label{U1abc}\end{equation}
We will see later how these emerge from the geometry.  For now, however, we note that $U(1)_a$ and $U(1)_b$ both prevent the generation of the operator $\int\,d^4\theta\,X^{\dag} H\bar{H}$ which we use to obtain $\mu$.  To get around this, we must adopt the philosophy of \cite{Beasley:2008kw} and assume that some dynamics on the $a$ and $b$ matter branes cause these fields to pick up nonzero expectation values.  In that case, both $U(1)_a$ and $U(1)_b$ are Higgs'ed and the $\mu$ term can be generated.  

The success of the setup in figure \ref{su7mu} depends largely on one's point of view.  On the one hand, it is disadvantageous relative to the case of $SO(12)$ intersections because we are forced to introduce new arbitrariness into the model regarding the dynamics of these new gauge singlets.  On the other hand, one could view this instead as an advantage because, from the bottom-up perspective, we can think of $a$ and $b$ as a pair of coupling constants which give us greater tunability.

In this paper, we would prefer to have models that are as complete as possible without introducing extra dynamics so in what follows we will devote most of our attention to the case of $SO(12)$ intersections.  Nevertheless, we find it very encouraging that the general scenario of section \ref{subsec:muBmu}, in which $X$ picks up a nonzero $U(1)_{PQ}$ charge, can emerge naturally regardless of how the Higgs and messenger curves intersect.  One possibility for the extra dynamics needed to give nonzero expectation values to the fields $a$ and $b$ is currently under investigation and will appear soon \cite{ourfuture}.

Finally, we note that as in the case of $SO(12)$ enhancements, the setup of figure \ref{su7mu} is in fact only one of two possibilities of this type.  The other, in which the $f$ and $\bar{f}$ curves are interchanged, still contains a coupling of $H$ and $\bar{H}$ to the SUSY-breaking sector via KK modes on the messenger curves.  The $PQ$ charge of $X$ is flipped in this setup, though, preventing generation of the operator $\int\,d^4\theta\,X^{\dag}H\bar{H}$ and hence forbidding $\mu$ entirely.

\subsubsection{$U(1)_{PQ}$ from Geometry}
\label{subsubsec:SU7U1PQ}

Finally, let us comment briefly on how the $U(1)$ charges \eqref{U1PQ2} and \eqref{U1abc} emerge from the geometry.  Again, we get a $U(1)$ from each matter brane.  The charges of the various fields under their matter brane $U(1)$'s are
\begin{equation}\begin{array}{c|c|c|c|c|c|c|c}
 & X & f & \bar{f} & a & b & H & \bar{H} \\ \hline
U(1)_f & -6 & 6 & 0 & 0 & -6 & 0 & 0 \\
U(1)_{\bar{f}} & 6 & 0 & -6 & 6 & 0 & 0 & 0 \\
U(1)_H & 0 & 0 & 0 & -6 & 0 & 6 & 0 \\
U(1)_{\bar{H}} & 0 & 0 & 0 & 0 & 6 & 0 & -6
\end{array}\end{equation}
Now, it is easy to see that the $U(1)$'s in \eqref{U1PQ2} and \eqref{U1abc} are given by
\begin{equation}\begin{split}Q_{PQ} &= \frac{1}{6}\left(Q_{\bar{f}} - Q_f + Q_H - Q_{\bar{H}}\right)\\
Q_a &= -\frac{Q_H}{6}\\
Q_b &= \frac{Q_{\bar{H}}}{6}\\
Q_c &= \frac{1}{6}\left(Q_f+Q_{\bar{f}}+Q_H+Q_{\bar{H}}\right)\,.
\end{split}\end{equation}


\section{Supersymmetry breaking and D3-instantons}
\label{sec:SusyBreaking}

We would now like to incorporate these ideas into "complete" models of gauge mediated supersymmetry breaking in full $F$-theory GUTs.  To do this, however, we first need to specify the dynamics which causes the spurion field $X$ to acquire an $F$-component expectation value.
As mentioned in section \ref{sec:GM}, our implementation of gauge mediation naturally realizes a setup of \cite{ourinst} in which D3-instantons trigger supersymmetry breaking via a Polonyi model.

\begin{figure}
\begin{center}
\epsfig{file=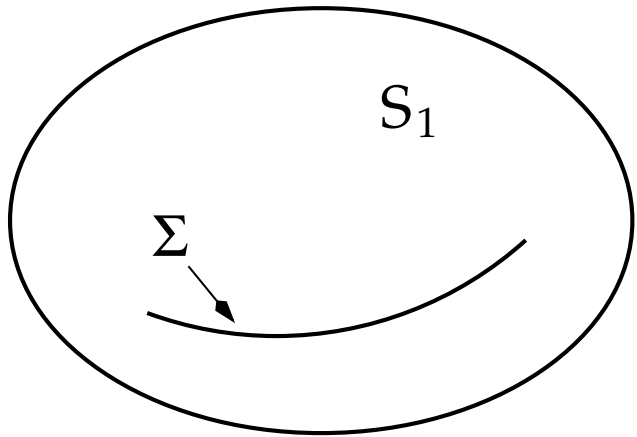,width=0.5\textwidth}
\label{polbranefig}
\caption{Basic Setup for Engineering Polonyi}
\end{center}
\end{figure}

\subsection{Setup}

The basic setup of the Polonyi model of \cite{ourinst} consists of a pair D7 branes wrapping 4-cycles $S_1$ and $S_2$, which
we choose to be del Pezzo surfaces, that intersect over a curve
$\Sigma$. In the following, we denote $S_1 = dP_M$ and $S_2 = dP_N$ and require
agreement of the canonical classes restricted to $\Sigma$
$$K_{S_1}|_{\Sigma}=K_{S_2}|_{\Sigma}$$
in order to avoid working with twisted gauge bundles on del Pezzos. To engineer chiral matter localized on $\Sigma$, we turn on nontrivial supersymmetric line bundles 
$V_a$ for the
$U(1)_a$ gauge fields along $S_a$. Recall that a supersymmetric bundle $V_a$ and $S_a$ must satisfy \be\label{SusyBundle} \int_{S} c_1(V_a)\wedge J^{(a)} =0 \,. \ee  In our local model, we will assume that the K\"ahler forms $J^{(a)}$ on $S_a$ are given by
 \be \label{kahler}
 \ba
J^{(1)} &=A^{(1)}H-\sum_{i=1}^M B^{(1)}_iE_i \cr
J^{(2)} &=A^{(2)}H'-\sum_{j=1}^N B^{(2)}_jE'_j \,,
\ea
\ee
where
\be
 A^{(a)} \gg 1\,, \qquad B^{(1)}_i\,, B^{(2)}_j \sim \mathcal{O}(1)\,, \hbox{ for all } i,j \,, 
 \ee 
and
${H,E_i}$ and ${H',E'_j}$ are bases of $H_2(S_1,\mathbb{Z})$ and
$H_2(S_2,\mathbb{Z})$, respectively. 
Further, as explained in \cite{ourinst},
we choose $\Sigma = \mathbb{P}^1$ and bundles on $S_i$ such that the
spectrum consists of only one chiral multiplet $X$ with charges
$(+, -)$.  For example, this is ensured for 
\be 
V_2|_{\Sigma} = V_1|_\Sigma
\otimes \mathcal{O}(-1)\,, 
\ee 
since then 
\be \ba n_{+-0}&=
h^0\left(\Sigma,K_{\Sigma}^{1/2}\otimes V_1|_{\Sigma}\otimes
V_2^{-1}|_{\Sigma}\right) =1\cr n_{-+0}&=
h^0\left(\Sigma,K_{\Sigma}^{1/2}\otimes V_1^{-1}|_{\Sigma}\otimes
V_2|_{\Sigma}\right)=0\,, 
\ea \ee
where $n_{pq0}$ denotes the number of multiplets of charges $(p,q)$ under $U(1)_1\times U(1)_2$.


\subsection{D3-instantons}

As shown in \cite{ourinst}, D3-instanton effects in this setup can generate a Polonyi model.
To see how the requisite superpotential term $W\sim X$ is generated, 
consider a D3-instanton wrapped on $S_1$ with gauge group $U(1)_{\text{inst}}$
and associated supersymmetric bundle $V_{\text{inst}}$.
The number of zero modes, $n_{pqr}$, from the D3-instanton to the D7's on $S_1$ and $S_2$, respectively,  with charges $(pqr)$ under $U(1)_{1} \times U(1)_2 \times U(1)_{\text{inst}}$ are then counted by
\be
\ba
n_{+0-}&= h^1\left(S_1,V_1\otimes V_{\text{inst}}^{-1}\right)=-\chi\left(S_1,V_1\otimes V_{\text{inst}}^{-1}\right)\cr
n_{-0+}&= h^1\left(S_1,V_1^{-1}\otimes V_{\text{inst}}\right)=-\chi\left(S_1,V_1^{-1}\otimes V_{\text{inst}}\right)\cr
n_{0+-}&= h^0\left(\Sigma,K_{\Sigma}^{1/2}\otimes V_2|_{\Sigma}\otimes V_{\text{inst}}^{-1}|_{\Sigma}\right)\cr
n_{0-+}&= h^0\left(\Sigma,K_{\Sigma}^{1/2}\otimes V_2|_{\Sigma}^{-1}\otimes V_{\text{inst}}|_{\Sigma}\right) \,.
\ea
\ee
In order to generate a linear term in the superpotential we require
\be
n_{-0+}=n_{0+-}=1\qquad\text{and}\qquad n_{+0-}=n_{0-+}=0 \,,
\ee
which will yield a coupling of the form $X \alpha_{0+-} \beta_{-0+}$ in the instanton action.  This in turn generates the required linear term in the  superpotential \cite{ourinst}
\be
W_{\text{inst}} = F_X X \,.
\ee
Here, $F_X$ is exponentially suppressed by the instanton action $e^{-t}$ as
\be
F_X \sim M_{Pol}^2 e^{-t} \,,
\ee
and $1/M_{Pol}$ is size of the 4-cycle $S_1$ that is wrapped by the instanton. 

Similarly, a D3-instanton wrapped on $S_2$  with a line bundle $V_{\text{inst'}}$ on its
world-volume generates a linear term if
\be
n'_{0+-}=n'_{-0+}=1\qquad\text{and}\qquad n'_{0-+}=n'_{+0-}=0 \,,
\ee
where  $n'_{pqr}$ counts modes with charges $(pqr)$ under $U(1)_{1} \times U(1)_2 \times U(1)_{\text{inst'}}$.  These, in turn, are determined by
\be
\ba
n'_{0+-}&= -\chi\left(S_2,V_2\otimes V_{\text{inst'}}^{-1}\right)\cr
n'_{0-+}&= -\chi\left(S_1,V_2^{-1}\otimes V_{\text{inst'}}\right)\cr
n'_{-0+}&= h^0\left(\Sigma,K_{\Sigma}^{1/2}\otimes V_1^{-1}|_{\Sigma}\otimes V_{\text{inst'}}|_{\Sigma}\right)\cr
n'_{+0-}&= h^0\left(\Sigma,K_{\Sigma}^{1/2}\otimes V_1|_{\Sigma}\otimes V_{\text{inst'}}^{-1}|_{\Sigma}\right) \,.
\ea
\ee


\subsection{Sum over instantons}
\label{sec:SumInst}

To obtain the complete superpotential, however, we need to sum over all possible instanton configurations and bundles $V_{\text{inst}}$.
Vital for supersymmetry breaking is that no higher order terms in $X$ are generated in this way.  Fortunately, this has been addressed in \cite{ourinst}.
Denote by $\{H, E_i\}$ the basis of $H_2(S_1, \mathbb{Z})$. For the ansatz
\be
\mathcal{L} = V_{\text{inst}} \otimes V_1^{-1} = b_0 H + \sum_{i=1}^M b_i E_i \,, \qquad b_j \in \mathbb{Z}\,,
\ee
it was found that a superpotential term of the form
\be
W_{\text{inst}}  \sim X^m \,, \qquad m\in \mathbb{N}_+
\ee
is generated only for ${\cal{L}}$ satisfying
\be\label{BundleConstraints}
\chi(S_1,\mathcal{L})=-m,\qquad
\chi(S_1,\mathcal{L}^{-1})=0,\qquad
\mathcal{L}\vert_{\Sigma}=\mathcal{O}(-m-1) \,.
\ee
For fixed intersection curve $\Sigma$ (of genus $0$ in the present case) we need to sum over all supersymmetric bundles $\mathcal{L}$ solving these constraints.
For $S_1 =dP_M$ with $M=3,\ldots,8$ and the class of $\Sigma$ in $H_2(S_1,\mathbb{Z})$ chosen as
\be
[\Sigma]=H-E_1-E_2 \,,
\ee
it was demonstrated in \cite{ourinst} that
there are no supersymmetric solutions of (\ref{BundleConstraints}) for $m>1.$ Meanwhile, all supersymmetric
solutions with $m=1$ have the form
\be
\label{PolBundles}
\mathcal{L}_p = \mathcal{O} (E_p -E_1 -E_2) \,,\qquad p=3,\cdots, M\,,
\ee

Each non-trivial solution contributes to a Polonyi linear superpotential for the chiral superfield $X${\footnote{As discussed in \cite{ourinst}, we must also sum over multi-instanton contributions.  In general, an $m$-instanton configuration can generate a superpotential coupling $X^m$ which is suppressed by a factor $e^{-mS_{\text{inst}}}$.  Such terms do not destabilize the SUSY-breaking vacuum, however, and are in fact completely negligible there.}}.

Similarly, to generate $X^m$ with $m\ge 1$ from a D3-instanton wrapped on $S_2$
one has to require
\be\label{BundleConstraintsii}
\chi(S_2,\mathcal{L}'^{-1})=-m,\qquad
\chi(S_2,\mathcal{L}')=0,\qquad
\mathcal{L}'\vert_{\Sigma}=\mathcal{O}(m+1) \,.
\ee
where $\mathcal{L}'=V_2^{-1}\otimes V_{\text{inst'}}.$
These equations can be obtained from (\ref{BundleConstraints})
if we replace $S_1$ with $S_2$ and $\mathcal{L}$ with $\mathcal{L}^{'\, -1}.$

It is easy to ensure that there is no contribution to the
superpotential arising from D3-instantons on $S_2$. For example, one
may consider $S_2=dP_2$  and choose the class of $\Sigma$ in
$H_2(S_2,\mathbb{Z})$ as $[\Sigma]'=H'-E'_1-E'_2.$ Then there are no
solutions of  (\ref{BundleConstraintsii}) for any $m\ge 1$.

Alternatively, we may choose $S_2=dP_N$ with $N=3,\ldots,8$ and
$[\Sigma]'=H'-E'_1-E'_2.$ Then, no higher terms $W\sim X^m$ with
$m>1$ are generated from a D3-instanton on $S_2.$ Meanwhile, one has
to sum over instanton bundles giving rise to linear terms $W\sim X$, i.e.
\be \label{bundlemore} V_{\text{inst}'}=V_2\otimes \mathcal{L}_p' \,,
\ee 
with
\be
\mathcal{L}_p'=\mathcal{O}(E'_1+E'_2-E'_p)\,,\qquad p\ne 1,\, p\ne 2 \,.
\ee


\section{A Complete Local Model}
\label{sec:completemodel}

We now turn our attention to the construction of complete models in which an explicit SUSY-breaking sector, such as the one discussed in section \ref{sec:SusyBreaking},
is coupled to an $SU(5)$ GUT model within the gauge mediation framework of sections \ref{sec:GM} and \ref{sec:HiggsMu}.  We shall proceed in two steps.  First, we shall discuss in more detail the natural emergence of the Polonyi model of \cite{ourinst} in the setup of section \ref{sec:GM} and how it couples to the messenger sector.  After that, we shall couple this system to one of the $SU(5)$ GUTs of BHV II \cite{Beasley:2008kw}.  The result will be a local GUT model with realistic matter content and an implementation of gauge-mediated supersymmetry breaking which addresses both the $\mu$ and $\mu/B_{\mu}$ problems in a natural way.


\subsection{Coupling Polonyi to an F-theory GUT}
\label{subsec:PolonyiGUT}

The first step in building a "complete" model is to provide a specific SUSY-breaking sector and describe how it couples to the messenger fields.  For us, this is easily achieved because the intersecting 7-branes used to introduce messenger fields in section \ref{sec:GM} are precisely what we needed to realize the D3-instanton triggered Polonyi superpotential described in \cite{ourinst} and reviewed in section \ref{sec:SusyBreaking}.  As such, our combined SUSY-breaking and messenger sectors have superpotential of the following simple form
\begin{equation}
W = F_X X + \lambda_X Xf\bar{f}\label{POGM}\,,
\end{equation}
where $F_X$ is exponentially suppressed by a factor of the D3 instanton action{\footnote{As discussed in \cite{ourinst}, $m$-instantons will also generate $X^m$ interactions.  These corrections will be parametrically small in all situations considered in this paper so we shall simply ignore them.  Note that one cannot simply scale away the instanton-generated prefactors by a field redefinition because they will reappear in the K\"ahler potential.}}.  

\subsubsection{Lifting the Flat Direction of Polonyi}

To study SUSY-breaking in more detail, let us consider first the model without messengers, $\lambda_X=0$.  Because this is a simple Polonyi model with a flat potential, an important role is played by nonrenormalizable operators generated by UV physics that have thus far been ignored.  For instance, we have an anomalous $U(1)$ in the problem which becomes massive via the Green-Schwarz mechanism.  Integrating out the massive vector multiplet yields a quartic correction to the K\"ahler potential of the form \cite{ArkaniHamed:1998nu}
\begin{equation}\delta K \sim -\frac{c (X^{\dag}X)^2}{M_{GB}^2} \,,
\end{equation}
with $c>0$, and where $M_{GB}$ is the gauge-boson mass.  
This correction favors a stable vacuum at $\langle X\rangle=0$.  Because the gauge boson mass arises from coupling to a closed string axion, though, the scale $M_{GB}$ is sensitive to details of moduli stabilization.  While the string scale seems like one natural estimate for $M_{GB}$ in perturbative string compactifications, it is known that much smaller values can also be obtained \cite{Ibanez:1998qp}.  In the present $F$-theory framework, this scale can be estimated, as in \cite{Beasley:2008dc}, by that of the flux responsible for inducing chirality into the spectrum
leading to 
$$M_{GB}=\Lambda,\qquad \Lambda=min(M_1^{KK},M_2^{KK}).$$ 
Recall that $X$ lives at the intersection of $S_1$ and $S_2$ and we let $M_a^{KK}$ be the KK scale for $S_a.$
($M_1^{KK}$ was previously called $M_{Pol}.$)

A second source of corrections arises from integrating out Kaluza-Klein (KK) modes on $S_1$ and $S_2$ and their intersection.  In order to systematically compute such corrections, one in principle needs to know detailed information about the spectrum of KK modes as well as their coupling to $X$.  However, we can learn something about the general structure by studying a truncated toy model.  Along these lines, one can easily demonstrate \cite{ourfuture} that including only the lightest KK modes, which directly couple to $X$ in the superpotential, yields precisely the simple O'Raifeartaigh model studied in \cite{Shih:2007av}.  Integrating out the KK modes in this model generates a Coleman-Weinberg potential that lifts the flat direction and produces a stable 
SUSY-breaking vacuum in the parameter regime of interest for us, $\Lambda^2\gg F_X$.  

We assume that $\Lambda$ is comparable\footnote{We will see later that this is consistent with  what follows from imposing $m_{3/2}\sim 1$ GeV.} with KK scale of $S_{GUT}$
$$M_{GUT}=\eta \Lambda,\qquad \eta \sim 1.$$ 
Therefore the physics at the GUT scale stabilizes the flat potential of our Polonyi model which
can be encapsulated by GUT-suppressed contributions to the K\"ahler potential
\begin{equation}\delta K\sim -\frac{a|X|^4}{M_{GUT}^2} + \frac{b|X|^6}{M_{GUT}^4}+\ldots \,,
\label{kahlcorrs}\end{equation}
with coefficients $a,b,\ldots$ leading to a stable vacuum at
\begin{equation}\langle X\rangle = M + \theta^2 F_X \,.
\end{equation}
As in the truncated model of \cite{Shih:2007av}, we expect that for a wide range of KK masses the quartic correction is generated with $a>0$, leading to a stable vacuum at $M=0$.  
\subsubsection{Shifting $M$ with Gravitational Effects}

Let us now bring the messengers back into the game by setting
$\lambda_X\ne 0$.  It may seem that we run into trouble when $M=0$ because this expectation value is responsible for providing a mass to the messenger fields, $f$ an $\bar{f}$.  In fact, coupling a model with SUSY-breaking vacuum at $M=0$ to messenger fields as in \eqref{WOGM} renders this vacuum unstable to a supersymmetric one with nonzero expectation values{\footnote{We thank C.~Vafa for emphasizing to us the importance of this point.}} for $f$ and $\bar{f}$.  

Fortunately, it has been noted in \cite{Kitano:2006wz} that the vacuum at $M=0$ can be shifted to a nonzero value when the Polonyi model with K\"ahler corrections \eqref{kahlcorrs}  is coupled to gravity{\footnote{We are very grateful to R.~Kitano for a number of enlightening discussions on this point. }}.  It might sound surprising at first that gravitational effects could have such an impact since we typically expect them to be Planck-suppressed and hence completely negligible at the energy scales under consideration.  However, because $F_X X$ is gauge invariant in the fundamental theory before we fix any of the moduli, we nevertheless expect on general grounds that the full potential (including gravity) will contain a linear term capable of inducing precisely such a shift, namely
\begin{equation}V\sim \ldots + \tilde{M}\left(F_X X + F_X^{\ast}X^{\dag}\right)+\ldots\,.
\label{linterm}\end{equation}
Because this term is absent in our field theoretic description, the dimensionful parameter $\tilde{M}$ will be Planck-suppressed.  Nevertheless, we must be careful before using this fact to simply throw it away because the Planck-suppressed contribution to this term, though small, is the \emph{leading} one.   

It is in fact easy to see how such a term can arise in our setup.  In general, the superpotential will contain contributions from sources away from the GUT stack, such as fluxes or additional 7-branes.  At sufficiently low energies, we can model this by adding a constant $W_0$ to the superpotential
\begin{equation}W\sim F_X X + W_0 \,.
\end{equation}
While this has no effect on the $M_{Pl}=\infty$ potential, it can play a role when $M_{Pl}$ is large but finite.  At energies smaller than $M_{GUT}$ where 4-dimensional SUGRA is reliable, for instance, one can see directly that $W_0$ modifies the SUGRA potential by adding precisely a linear term of the sort \eqref{linterm}
\begin{equation}
V_{SUGRA}\sim \frac{1}{M_{Pl}^2}\left(W_0^{\ast}F_XX+W_0F_X^{\ast}X^{\dag}\right) + \frac{a|F_X|^2|X|^2}{M_{GUT}^2}+\ldots\,.
\label{VSUGRA}\end{equation} 
In the presence of $W_0$, then, the vacuum at $M=0$ is shifted to
\begin{equation}M \sim \frac{|W_0| M_{GUT}^2}{|F_X| M_{Pl}^2} \,.\end{equation}
Though we do not know the value of $W_0$ from first principles, we can obtain a reasonable estimate by following the suggestion of \cite{Kitano:2006wz} and imposing the constraint that $V\sim 0$ at the vacuum.  This leads to $|W_0|\sim |F_X| M_{Pl}$ and hence to the estimate
\begin{equation}M\sim \frac{M_{GUT}^2}{M_{Pl}}\,,
\end{equation}
which we will use throughout the rest of this paper.
If we take a strict limit $M_{Pl}\rightarrow\infty$ with $M_{GUT}$ fixed then we recover $M=0$ as expected.  However, $M_{GUT}^2/M_{Pl}$ is in reality around $10^{14}$ GeV, a scale which is small in Planck units but nevertheless gives a sizeable mass to the messenger fields and is sufficiently far from the origin that this vacuum can remain metastable and long-lived when the messengers are included \cite{Kitano:2006wz}.



\subsection{Coupling to a BHV $SU(5)$ GUT}

To construct a complete model, we consider a slight modification of the $SU(5)$ Model II of BHV II which we will call Model II'.  We take the hypercharge bundle ${\cal{L}}_Y={\cal{O}}(E_3-E_4)^{1/5}$ as in BHV II and consider matter branes intersecting the GUT brane along the curves indicated in the table below,
where all the entries are taken from equation (17.9) of \cite{Beasley:2008kw} except the second row.

\begin{center}
\begin{tabular}{|l|l|l|l|l|l|} \hline
Model II' & Curve & Class & $g_{\Sigma}$ & $L_{\Sigma}$ & $L_{\Sigma}^{\prime\,n}$ \\ \hline
$1\times 5_H$ & $\Sigma_H^{(u)}$ & $H-E_1-E_3$ & 0 & ${\cal{O}}_{\Sigma_H^{(u)}}(1)^{1/5}$ & ${\cal{O}}_{\Sigma_H^{(u)}}(1)^{2/5}$ \\ \hline
$1\times \bar{5}_H$ & $\Sigma_H^{(d)}$ & $H - E_4 - E_5$ & 0 & ${\cal{O}}_{\Sigma_H^{(d)}}(-1)^{1/5}$ & ${\cal{O}}_{\Sigma_H^{(d)}}(1)^{2/5}$\\ \hline
$3\times 10_M$ & $\Sigma_M^{(1)}$ (pinched) & $2H - E_1 - E_5$ & 0 & ${\cal{O}}_{\Sigma_M^{(1)}}$ & ${\cal{O}}_{\Sigma_M^{(1)}}(3)$ \\ \hline
$3\times\bar{5}_M$ & $\Sigma_M^{(2)}$ & $H$ & 0 & ${\cal{O}}_{\Sigma_M^{(2)}}$ & ${\cal{O}}_{\Sigma_M^{(2)}}(3)$ \\ \hline
\end{tabular}
\end{center}

Here, $g_{\Sigma}$ is the genus of a given matter curve, $\Sigma$, and $L_{\Sigma}$ is the restriction of the hypercharge bundle to $\Sigma$.  In addition, $L_{\Sigma}'$ is the restriction of the bundle on the corresponding matter brane to its matter curve $\Sigma$ and $n$ is the charge of the field in question with respect to the $U(1)$ on the matter brane. For example, $n=6$ for $5_H$ and $n=-6$ for ${\bar 5}_H.$
Note that the bundles  $L_{\Sigma_{H^{(u)}}}$ and $L_{\Sigma_{H^{(u)}}}^{\prime}$ are chosen in such
a way that only the Higgs doublet $H^{(u)}$ remains massless on ${\Sigma_{H^{(u)}}}.$
Similarly, only the Higgs doublet $H^{(d)}$ is massless on ${\Sigma_{H^{(d)}}}.$
Nevertheless, we keep the notations of BHV II in the left-most column of the table
in order to simplify the presentation
of superpotential couplings by emphasizing the  origin of these fields as coming from
$5_H$ and ${\bar 5}_H$ respectively.

In the above table, we have made only one modification
to the $SU(5)$ Model II of BHV II \cite{Beasley:2008kw} and that is to
change the class of the $\bar{5}_H$ matter curve from $H-E_1-E_3$ to
$H-E_4-E_5$. Indeed with the choice as in BHV II $L_{\Sigma_{H^{(d)}}}=\mathcal{O}(1)^{1/5}$ and  leads to
a doublet on this curve which comes from  $5_H$ rather than  $5_{\bar{H}}$.

Note that with our choice, the intersection
number of $\Sigma_{H^{(u)}}$ with $\Sigma_{H^{(d)}}$ is still non-zero so these matter curves will intersect one another in general.  This could pose a problem for the general program of section \ref{sec:HiggsMu} because $\mu$ could be generated even before the messenger sector is added.  We therefore digress for a moment to discuss the nature of these intersections in a bit more detail.

\subsubsection{$\mu$ from Intersection of Higgs Matter Curves?}

As with intersections of Higgs and messenger curves in section \ref{sec:HiggsMu}, the $H$ and $\bar{H}$ curves are regions of $SU(6)$ enhancement and can meet at points with either an $SU(7)$ or an $SO(12)$ enhancement.  In the case of $SO(12)$ enhancement, a third matter curve which engineers a $\mathbf{10}$ must emanate from the singular point.   We would prefer to avoid adding extra matter curves or including extra intersections with the $\mathbf{10}$ matter curves already present so we suppose that if the $H$ and $\bar{H}$ matter curves meet then the singularity at the intersection point is $SU(7)$.  

The case of $SU(7)$ enhancement has been discussed in \cite{Beasley:2008kw} and leads to couplings $\lambda_{\psi_i}\psi_i H\bar{H}$ for some GUT singlet fields $\psi_i$.  If any $\psi_i$ picks up a nonzero expectation value then this generates a $\mu$ term which can be small if the corresponding $\lambda_{\psi_i}$ is suppressed due to repulsion of the $\psi_i$ wave function from the GUT brane \cite{Beasley:2008kw}.  In order to connect the generation of $\mu$ to SUSY-breaking as in section \ref{sec:HiggsMu}, we would like to avoid this scenario so if the $H$ and $\bar{H}$ curves intersect at an $SU(7)$ enhanced point we prefer all of the $\psi_i$ to have vanishing expectation values.  One easy way to achieve this is to choose bundles on the $H$ and $\bar{H}$ matter branes so that none of the $\psi_i$ are zero modes.  In that case, we expect the large KK scale masses to drive their expectation values to zero{\footnote{One might worry that a superpotential term of the form $W\sim \psi$ could lead to a nonzero expectation value.  We are only aware of one way such a term could be generated without adding anything further to this construction and that is via a D3-instanton.  As discussed in \cite{ourfuture}, though, D3-instantons wrapping Higgs matter branes cannot generate such couplings due to the presence of extra 3-7 and 7-3 zero modes connecting the instanton to the GUT brane.}}.

\subsubsection{Adding the SUSY-Breaking Sector}

We now add in our SUSY-breaking sector.  To do this, we need only specify the $f$, $\bar{f}$, and $\phi$ matter curves.  This is summarized in the following table:

\begin{center}
\begin{tabular}{|l|l|l|l|l|l|} \hline
 & Curve & Class & $g_{\Sigma}$ & $L_{\Sigma}$ & $L_{\Sigma}^{\prime\,n}$ \\ \hline
$1\times 5_f$ & $\Sigma_f$ & $E_1$ & 0 & ${\cal{O}}_{\Sigma_f}$ & ${\cal{O}}_{\Sigma_f}(1)$  \\ \hline
$1\times \bar{5}_{\bar{f}}$ & $\Sigma_{\bar{f}}$ & $H-E_1-E_6$ & 0 & ${\cal{O}}_{\Sigma_{\bar{f}}}$ & ${\cal{O}}_{\Sigma_{\bar{f}}}(1)$ \\ \hline
$KK \times \left(10_{\phi}+\bar{10}_{\bar \phi}\right)$ & $\Sigma_{\phi}$ & $2H-E_1-E_2-E_5$ & 0 & ${\cal{O}}_{\Sigma_{\phi}}$ & ${\cal{O}}_{\Sigma_{\phi}}$ \\
\hline
\end{tabular}
\end{center}

Note that  intersection numbers of $\Sigma_f$, $\Sigma_{\bar{f}}$, and $\Sigma_{\phi}$ with each other and with $\Sigma_{H^{(u)}}$ and $\Sigma_{H^{(d)}}$ are consistent with the intersections that we drew in figure \ref{gettingmu}.  These triple intersections satisfy the consistency conditions spelled out in \cite{Beasley:2008kw} so this choice of curves effectively implements our gauge mediated supersymmetry-breaking scenario in this particular $F$-theory GUT.

Note that this situation is not completely optimal because our
messenger curves $\Sigma_f,\Sigma_{\bar{f}},\Sigma_{\phi}$ will also
intersect the matter curves $\Sigma_M^{(1)}$ and $\Sigma_M^{(2)}$.
It is easy to arrange these intersections so that the only new superpotential interactions involve KK modes{\footnote{One way to accomplish this is as follows.  First let $\Sigma_f$ be a pinched curve which meets $\Sigma_M^{(1)}$ at an $SO(12)$ point.  Then let $\Sigma_{\overline{f}}$ meet $\Sigma_M^{(2)}$ at an $SU(7)$ point with bundles chosen so that there are no GUT singlet zero modes there.  Further, let $\Sigma_{\overline{f}}$ meet $\Sigma_M^{(1)}$, which we require to have a second pinching, at an $E_6$ point.  Finally, let $\Sigma_{\phi}$ meet $\Sigma_M^{(1)}$ and $\Sigma_M^{(2)}$ at an $E_6$ point.  These intersections yield four new types of interaction but each one necessarily involves KK modes.}}.
Nevertheless, they will give rise to KK-suppressed superpotential couplings and also possibly $D$-term couplings which could in principle be problematic for phenomenology.


\subsubsection{Polonyi in Model II'}
\label{subsec:PolonyiModel}

While we have specified cycles and bundles on the GUT brane, it remains to discuss analagous details of the Polonyi model specific to this construction.
Recall that, in this model, the intersecting 7-branes of section \ref{sec:SusyBreaking} are the matter branes $S_f$ and $S_{\bar{f}}$ which intersect the GUT brane stack along the curves $\Sigma_f$ and $\Sigma_{\bar{f}}$, respectively.  As in section \ref{sec:SusyBreaking}, we require $S_f$ and $S_{\bar{f}}$ to intersect along a curve $\Sigma_{Pol}$.  The curve $\Sigma_{Pol}$, in turn, meets the GUT brane at a single point of $SU(7)$ enhancement.

Before describing specific conditions for the bundles $V_f,V_{\bar{f}}$ on $S_f,S_{\bar{f}}$, we must first note that the normalization of the $U(1)$'s on $S_f$ and $S_{\bar{f}}$ in \cite{Beasley:2008kw} and hence the tables which define our current model are derived from Slansky's conventions \cite{Slansky:1981yr} for the decomposition $SU(7)\rightarrow SU(5)\times U(1)\times U(1)$ and hence differ from those in section \ref{sec:SusyBreaking}.  To make a connection with the results of section \ref{sec:SusyBreaking}, then, we note that the bundles $V_1$ and $V_2$ contained therein are related to $V_f$ and $V_{\bar{f}}$ by
\begin{equation}
V_1=V_f^6 \,,\qquad V_2=V_{\bar{f}}^6\,.
\end{equation}

We now describe the conditions for obtaining a Polonyi superpotential from a D3-instanton wrapping $S_f$.  As discussed in section \ref{sec:SumInst} we take $S_f=dP_M$ for $3\le M\le 8$ and $S_{\bar{f}}=dP_2$.  We must also specify the class of $\Sigma_{Pol}$ in $S_f$ and $S_{\bar{f}}$.  Denoting the former by $\tilde{\Sigma}$ and the latter by $\hat{\Sigma}$, we take{\footnote{Note that the classes of $\Sigma_{Pol}$ are chosen in such a way that $K_{S_f}|_{\Sigma_{Pol}}=K_{S_{\bar{f}}}|_{\Sigma_{Pol}}$ which allows us to avoid having to work with twisted gauge bundles on del Pezzos.}}
\begin{equation}
\tilde{\Sigma}_{Pol}=\tilde{H}-\tilde{E}_1-\tilde{E}_2,\qquad \hat{\Sigma}_{Pol}=\hat{H}-\hat{E}_1-\hat{E}_2\,,
\end{equation}
where $\tilde{H},\tilde{E}_i$ is the standard basis of $H_2(S_f,\mathbb{Z})$ and $\hat{H},\hat{E}_j$ the standard basis of $H_2(S_{\bar{f}},\mathbb{Z})$.

In order to have a single chiral field $X$ localized on $\Sigma_{Pol}$ which can couple to the combination $f\bar{f}$ we need
\begin{equation}
V_{\bar f}^6|_{\Sigma_{Pol}}=V_{f}^6|_{\Sigma_{Pol}}\otimes {\cal{O}}(-1)\,.
\end{equation}
Given this, we know from section \ref{sec:SusyBreaking} that, for a suitable choice of K\"ahler form, the only BPS D3-instantons on $S_f=dP_M$ which generate superpotential terms in the 1-instanton sector are
\begin{equation}
V_{\text{inst}}^6=V_f^6\otimes {\cal{L}}^{(k)}\,,
\end{equation}
where
\begin{equation}
{\cal{L}}^{(k)}={\cal{O}}\left(\tilde{E}_k-\tilde{E}_1-\tilde{E}_2\right),\qquad k=3,\ldots,M\,.
\label{Lksol}\end{equation}
Moreover, these will generate a Polonyi superpotential for $X$ provided there are no extra fermion zero modes between the D3-instanton and the GUT stack.  This is because the presence of such fermi zero modes will in general cause the superpotential contributions from the ${\cal{L}}^{(k)}$ to vanish.  
Since the restriction of the hypercharge bundle on the GUT stack to $\Sigma_f$, $L_{\Sigma_f}$, is trivial the D3-GUT fermion zero modes are counted as
\be \ba {\bf
5}_{+6} :&\qquad  h^0(\mathbb{P}^1,\mathcal{O}(-1)\otimes
V_{\text{inst}}|_{\Sigma_{f}}^6) \cr \bar{\bf 5}_{-6}: &\qquad
h^0(\mathbb{P}^1,\mathcal{O}(-1)\otimes
V_{\text{inst}}|_{\Sigma_{f}}^{-6})\,. \ea \ee We will choose the class of $\Sigma_f$ in $S_f$ in such a way that for at least one ${\cal{L}}^{(k)}$, the corresponding instanton bundle restricts trivially to $\Sigma_f$ 
\be \label{triv}
V_{\text{inst}}|_{\Sigma_{f}}=\mathcal{O},\ee so that there are no
fermion zero modes between the D3-instanton and the GUT D7-branes.  Looking at our table, we recall that $V_f^6|_{\Sigma_f}={\cal{O}}(1)$.  Combining this with \eqref{Lksol}, we see that the condition \eqref{triv} is equivalent to requiring that some ${\cal{L}}^{(k)}$ satisfies
\begin{equation}
{\cal{L}}^{(k)}|_{\Sigma_f}={\cal{O}}(-1)\,.
\end{equation}
This can be accomplished for one choice of $k$, for example if we take $M=7$ and 
\begin{equation}
\Sigma_f = 2{\tilde H}-\tilde{E}_1-{\tilde E}_4-{\tilde E}_5-{\tilde E}_6-{\tilde E}_7\,.
\end{equation}
In this case, we get a Polonyi superpotential from ${\cal{L}}^{(3)}$ and vanishing contributions from the remaining ${\cal{L}}^{(k)}$ with $k=4,\ldots,7$.

Finally, we note that $S_f$ may in general intersect other matter
branes besides the GUT brane and $S_{\bar f}$.  We therefore require
that these intersections occur only over $\mathbb{P}^1$'s and  the
restriction of the gauge bundles on $S_f$ and other matter branes to
any of these intersections is trivial so that we do not get any new
charged matter that would participate in instanton-induced
interactions. We also require that the instanton bundle $V_{inst}$
restricts trivially to these  $\mathbb{P}^1$'s so that there are no
extra fermion zero modes between the instanton and the other matter
branes.

In the above discussion only the restriction of the bundle $V_f$
to $\Sigma_{Pol}$ and $\Sigma_f$ was specified. However, it is easy to find
a supersymetric bundle $V_f$ on $S_f$ which  has such restrictions.

\section{Sweet Spot Supersymmetry from F-Theory}
\label{sec:numerology}

In this section, we turn our attention to the effective action of models of the type considered in section \ref{sec:completemodel} in which the gauge mediation setup of section \ref{sec:HiggsMu} is combined with the D3-instanton triggered Polonyi model of \cite{ourinst}, which was reviewed in section \ref{sec:SusyBreaking}.  In particular, we demonstrate that this setup provides a natural realization of Ibe and Kitano's "sweet spot supersymmetry" \cite{Ibe:2007km}.  The effective action of this model contains a number of dimensionless parameters as well as two dimensionful ones, which we can think of as the gravitino mass and the scale of new physics couplings the Higgs and messenger sectors.  For us, the latter dimensionful parameter is fixed to be the GUT scale but the remaining parameters remain unspecified and, because we are only discussing local models, a direct calculation of them is beyond the scope of this paper. 
Nevertheless, we now review the values required to reproduce the successful phenomenology of \cite{Ibe:2007km} and the degree to which they may be plausible in such constructions. 
Building compact toy models of the gauge mediation scenario in this paper where such claims could be directly tested would of course be of much interest.
 
Let us begin by recalling the form of the effective action for the messenger and Higgs sectors.  Combining the Polonyi superpotential \eqref{POGM} with the quartic corrected K\"ahler potential \eqref{kahlcorrs} and the higher dimension operators \eqref{highdimops} which are generated by integrating out KK modes, we obtain
\begin{equation}\begin{split}{\cal{L}} &\sim \int\,d^4\theta\,\left(X^{\dag}X - \frac{a(X^{\dag}X)^2}{M_{GUT}^2}+\frac{c_{\mu}X^{\dag}H\overline{H}}{M_{GUT}} + \frac{c_HX^{\dag}X\left(HH^{\dag}+\overline{H}\overline{H}^{\dag}\right)}{M_{GUT}^2}+\ldots\right)\\
&\qquad + \int\,d^2\theta\,\left(F_X X + \lambda_X X f\bar{f} + \frac{\tilde{\lambda}}{M_{GUT}}H\overline{H}f\overline{f}+\ldots\right)\,.
\label{oursweetspot}\end{split}\end{equation}
As discussed in section \ref{subsec:SO12}, the effect of the $\tilde{\lambda}$ coupling in \eqref{oursweetspot} is to yield an additional contribution to $c_{\mu}$ upon integrating out the massive messenger fields.  In the end, this leaves us with an effective action of precisely the sort studied in \cite{Ibe:2007km}, where this scenario was given the name "sweet spot supersymmetry".  The precise effective action of \cite{Ibe:2007km}, however, doesn't specify the suppression of higher dimension operators linking the Higgs and messenger sectors but rather replaces all appearances of $M_{GUT}$ in \eqref{oursweetspot} with an arbitrary scale $\Lambda$.  As such, they studied a model which depends on two dimensionful parameters, the arbitrary scale of new physics, $\Lambda$, and the gravitino mass, $m_{3/2}$.  It is quite remarkable that, when various phenomenological constraints were imposed on this effective action, Ibe and Kitano were led to conclude that $\Lambda$ had to be around $10^{16}$ GeV$\sim M_{GUT}$.  To be clear, the emergence of the GUT scale was a consequence of their analysis, not an input.  For us, on the other hand, there was never a choice for this scale.  Rather, we were forced to land on a model with a specific value of $\Lambda$ which fortunately seems to coincide with the phenomenologically preferred one.
 In order for the effective action of \eqref{oursweetspot} to be truly successful, though, we need a few additional conditions to hold, namely
\begin{itemize}
\item The OGM coupling $\lambda_X$ must be of ${\cal{O}}(10^{-2})$ or smaller
\item All dimensionless couplings which arise from integrating out KK modes must be of ${\cal{O}}(1)$
\item $m_{3/2}\sim 1$ GeV
\end{itemize}

The condition on $\lambda_X$ is necessary to ensure that the SUSY-breaking vacuum at $M\sim M_{GUT}^2/M_{Pl}$ remains stable when coupled to the messenger fields $f$ and $\bar{f}$ \cite{Kitano:2006wz}.  As discussed at length in BHV II \cite{Beasley:2008kw}, Yukawa couplings involving fields whose matter curves have size set by the GUT scale include a factor of $\alpha_{GUT}^{3/4}\sim 10^{-1}$ which then multiplies a wave function overlap integral.  This means that, to the extent that Yukawa couplings can have "natural" values, a rough estimate for $\lambda_X$ should be around $10^{-1}$ or so.  Note, however, that $\lambda_X$ can be further suppressed if the wave function of the GUT singlet field $X$ is repelled from the GUT brane \cite{Beasley:2008kw}.  

As for the dimensionless couplings $c_H$ and $c_{\mu}$, they encapsulate the effects of the full spectrum of KK modes so values of ${\cal{O}}(1)$ seem quite natural. As stressed in \cite{Ibe:2007km}, this is the same idea behind the Giudice-Masiero mechanism in gravity mediated models \cite{Giudice:1988yz}.  The only difference here is that the new physics comes in at the slightly lower scale $M_{GUT}$.  Note, however, that if $c_H$ and $c_{\mu}$ are obtained by perturbative loop integrals involving a small number of 4-dimensional massive fields then each will typically contain loop suppression factors involving the product of $1/16\pi^2$ and some number of Yukawa couplings{\footnote{Apart from detailed phenomenology, such loop factors can already cause a problem for naturalness of electroweak symmetry breaking, which requires $\mu^2\sim m_H^2$ and hence $c_{\mu}^2\sim c_H$.  This is reminiscent of the $\mu/B_{\mu}$ problem \cite{Dvali:1996cu}.}}.  To achieve ${\cal{O}}(1)$ coefficients, then, the Yukawas must be large enough to effectively cancel the $1/16\pi^2$.  This led \cite{Ibe:2007km} to suggest strongly coupled UV completions of the sort described in \cite{Kitano:2006wm}.  In our situation, the theory is in fact 8-dimensional at the GUT scale with matter fields localized on codimension 2 defects so estimates based on integrating out a few massive 4-dimensional fields do not obviously apply.  An honest computation of $c_H$ and $c_{\mu}$ in this context would be interesting but requires a more detailed knowledge of the various couplings in this 8-dimensional theory which is beyond the scope of this paper.

We finally turn to the gravitino mass $m_{3/2}$, which is determined by the instanton-generated scale $F_X$
\begin{equation}
m_{3/2}\sim \frac{F_X}{M_{Pl}}\,.
\end{equation}
In order to land on the $m_{3/2}\sim 1$ GeV "sweet spot", we need the instanton-generated scale $F_X$ to be of order $10^{19}$ GeV$^2$ or so.
The value of $F_X$ depends crucially on the size of the 4-cycle that the instanton wraps, though.  A natural choice for this size is $M_{GUT}^{-1}$ but let us include a possible ${\cal{O}}(1)$ deviation from this and write instead
\begin{equation}
M_{GUT}\sim \eta M_{Pol}\,,
\end{equation}
where $M_{Pol}^{-1}$ is the size of the cycle wrapped by the D3-instanton which generates our Polonyi model.  We also make an assumption regarding the tension of the instanton, namely that the relevant scale, $M_{\ast}$, is the same one which determines the tension of the GUT branes.
 This object, in turn, is related to the coupling constant at the GUT scale by\cite{Beasley:2008kw}
\begin{equation}\frac{M_{\ast}^4}{M_{GUT}^4}\sim 2\pi\alpha_{GUT}^{-1}\,.
\end{equation}

With all of these considerations, we can estimate $F_X$ as{\footnote{One subtlety that we overlook here is the effect of the anomalous $U(1)$ on this estimate.  In general, the coupling of its vector multiplet to closed string axions generates both an explicit mass $m$ as well as a Fayet-Iliopolous (FI) parameter $\xi$.  Treating them as fixed quantities, the FI parameter can be absorbed by a field redefinition but the ratio $\xi/m^2$ then appears in terms which violate this $U(1)$, such as as the superpotential coupling $F_X X$.  Unless $\xi/m^2$ is unusually large, this effect is not important.}}  
\begin{equation}
F_X\sim M_{Pol}^2\exp\left(-\frac{M_{\ast}^4}{M_{Pol}^4}\right)\sim \frac{M_{GUT}^2}{\eta^2}\exp\left(-2\pi\alpha_{GUT}^{-1}\eta^4\right)\,.
\end{equation}
Note that this result exhibits a very strong dependence on the ${\cal{O}}(1)$ number $\eta$ that we cannot compute from first principles without a compact model in hand.  What we can do, however, is demonstrate that an ${\cal{O}}(1)$ choice for $\eta$ can yield $F_X\sim 10^{19}$ GeV$^2$.  Indeed, this can be accomplished for
\begin{equation}\eta\sim 0.68\,,\end{equation}
which is not too far from unity{\footnote{Because our naive estimate for the anomalous $U(1)_{PQ}$ gauge boson mass is $\Lambda\sim M_{Pol}$, this is consistent with our earlier assumption that $\Lambda > M_{GUT}$.}}.  Small deviations of $\eta$ from this value, though, allow for a wide range of soft parameters so we can by no means "predict" the gravitino mass in this setup.  What we can say, however, is that this type of model is not obviously inconsistent with a 1 GeV gravitino.  



\section{Concluding Remarks}
\label{sec:conclusion}

In this paper, we have argued that gauge mediation can be easily incorporated into the $F$-theory GUTs of BHV \cite{Beasley:2008dc,Beasley:2008kw} and, in so doing, simple models of SUSY-breaking triggered by D3-instantons naturally appear.  Moreover, very naive couplings between the messenger and Higgs sectors lead to the emergence of a $U(1)_{PQ}$ symmetry with respect to which the SUSY-breaking spurion field $X$ is charged.  This connects the breaking of $U(1)_{PQ}$ to the breaking of SUSY and provides natural solutions to the $\mu$, $\mu/B_{\mu}$, and supersymmetric CP problems.  Moreover, it leads to an effective theory below the GUT scale of the sort that appears in phenomenologically viable models of "sweet spot supersymmetry" \cite{Ibe:2007km}.

While this gives us hope for realizing a successful model in the $F$-theory framework, much work remains to be done.  Of paramount importance is the successful embedding of local models of this type into full compactifications wherein various input parameters could, at least in theory, be determined from first principles.  Of course, this is quite an ambitious task for the fairly intricate collections of intersecting 7-branes that have so far been used to realize exotic-free GUT models in \cite{Beasley:2008kw} but even embeddings of simpler toy models of gauge mediated models in this framework would be desirable.

It is also important to develop a better understanding of the various superpotential couplings as it could allow for a more quantitative description of the flavor structure of F-theory GUTs.  In principle, one should be able to address this issue even in the local context, though certain assumptions about physics along noncompact directions may have to be made.
Early signs of progress along these lines have already appeared in \cite{Beasley:2008kw}, where the existence of a heavy generation was translated into a geometric condition.  Explicit computation of Yukawa couplings has also been achieved in some specific examples in different but related contexts \cite{Cremades:2003qj, Cvetic:2003ch, Cremades:2004wa,Braun:2006me,
Abe:2008fi,Conlon:2008qi}

\section*{Acknowledgements}

We would like to thank M.~Ibe, S.~Kachru, C.~Kilic, Y.~Ookouchi, H. Ooguri, K. Saraikin, K.~Vyas, T.~Weigand,
and especially C.~Beasley, J.~Heckman, R.~Kitano, and C.~Vafa for valuable discussions.
The work of JM and SSN was supported by John A. McCone Postdoctoral Fellowships.
The work of NS was supported in part by the DOE-grant DE-FG03-92-ER40701.
 JM would like to thank the SITP at Stanford University, the Simons Center for Geometry and Physics at Stony Brook, the 2008 Simons Workshop in Mathematics and Physics, and the Aspen Center for Physics for kind hospitality during the course of this work.
SSN thanks the IPMU and the University of Tokyo for kind hospitality.

\newpage


\bibliographystyle{JHEP}
\renewcommand{\refname}{Bibliography}

\addcontentsline{toc}{section}{Bibliography}


\providecommand{\href}[2]{#2}\begingroup\raggedright\endgroup

\end{document}